\documentclass[journal]{IEEEtran}

\usepackage{cite}				%引用
\usepackage{graphicx}	%图片环境
\usepackage{amsmath}		%数学环境
\usepackage{algorithmic}	%算法内部
\usepackage{array}		%
\ifCLASSOPTIONcompsoc	%子图标题设置
\usepackage[caption=false,font=normalsize,labelfont=sf,textfont=sf]{subfig}
\else
\usepackage[caption=false,font=footnotesize]{subfig}
\fi
\usepackage{stfloats}
\usepackage{url}
\usepackage{algorithm}	%算法浮动环境
\usepackage{color}		%字体颜色
\usepackage{bm}			%字体加粗
\usepackage{amsfonts}	%数学特殊字体包
\usepackage{amssymb}	%数学符号包
\usepackage[switch]{lineno}	%显示行号，附带\linenumbers 命令

\newtheorem{theorem}{Theorem}
\newtheorem{lemma}{Lemma}

\newtheorem{remark}{Remark}

\newtheorem{problem}{Problem}
\begin{document}
	%\linenumbers
	\title{Probabilistic Placement Optimization for Non-coherent and Coherent Joint Transmission in Cache-Enabled Cellular Networks}
	\author{Tianming~Feng, Shuo~Shi, Shushi~Gu, Wei~Xiang, and Xuemai~Gu, 
	\thanks{T. Feng, S. Shi, and X. Gu are with the School of Electronic and Information Engineering, Harbin Institute of Technology, Harbin 150001, China (e-mail: \{fengtianming, crcss, guxuemai\}@hit.edu.cn).}
	\thanks{S. Gu is with the School of Electronic and Information Engineering, Harbin Institute of Technology (Shenzhen), Shenzhen~518055, China (e-mail: gushushi@hit.edu.cn).}
	\thanks{W. Xiang is with the College of Science and Engineering, James~Cook~University, Cairns, QLD~4878, Australia (e-mail: wei.xiang@jcu.edu.au).}
	\vspace{-2em}}% <-this % stops a space
	\maketitle
	\begin{abstract}
		How to design proper content placement strategies is one of the major areas of interest in cache-enabled cellular networks. In this paper, we study the probabilistic content placement optimization of base station (BS) caching with cooperative transmission in the downlink of cellular networks. With placement probability vector being the design parameter, non-coherent joint transmission (NC-JT) and coherent joint transmission (C-JT) schemes are investigated according to whether channel state information (CSI) is available. Using stochastic geometry, we derive an integral expression for the successful transmission probability (STP) in NC-JT scheme, and present an upper bound and a tight approximation for the STP of the C-JT scheme. Next, we maximize the STP in NC-JT and the approximation of STP in C-JT by optimizing the placement probability vector, respectively. An algorithm is proposed and applied to both optimization problems. By utilizing some properties of the STP, we obtain globally optimal solutions in certain cases. Moreover, locally optimal solutions in general cases are obtained by using the interior point method. Finally, numerical results show the optimized placement strategy achieves significant gains in STP over several comparative baselines both in NC-JT and C-JT. The optimal STP in C-JT outperforms the one in NC-JT, indicating the benefits of knowing CSI in cooperative transmission.
	\end{abstract}
	
	\begin{IEEEkeywords}
		Cache-enabled cellular networks, joint transmission, optimization, stochastic geometry, CSI, STP.
	\end{IEEEkeywords}
	
	\IEEEpeerreviewmaketitle
	
\section{Introduction}
	
	\IEEEPARstart{D}UE to the rapid development of the Internet-of-Things (IoT) and the intellectualized evolution of mobile devices, mobile data traffic is growing exponentially in recent years, leading to an enormous burden on the current wireless network. According to the report from Cisco \cite{001Cisco}, global mobile data traffic will dramatically increase seven-fold in 2022 compared with that of 2017, and will reach 77.5 exabytes per month.
	Investigation reveals a large portion of mobile data traffic is generated by repeatedly downloading popular contents and applications (e.g., movies and videos) by different users. With the development of content delivery networks (CDNs), caching popular contents at the edge of networks has been considered as a promising approach to significantly alleviate network traffic load by utilizing idle storage resources in small base stations (BSs) or smart devices \cite{013Wang2014CacheIT,014MaddahAli2014FundamentalLO}. Using this technique, user demands for cached contents can be satisfied directly by edge devices without retrieving from core networks \cite{013Wang2014CacheIT}, and transmission reliability can be improved thanks to file diversity gains \cite{014MaddahAli2014FundamentalLO}.
	
	There are two main categories of research problems in caching-enabled cellular networks, i.e., content placement strategy \cite{22Batug2014CacheenabledSC,24Bharath2016ALA,23TamoorulHassan2015ModelingAA,12Cui2015AnalysisAO} and content delivery strategy \cite{015Cui2016AnalysisAO,Xing2017TemporalSpatialRA,018jiang2018partition,09DongDongjiangEnMultiAntenna}. Carefully designing the placement strategy can take full advantage of file diversity gains provided by caching due to the limited size of the local cache. Specifically, one can cache the most popular contents \cite{22Batug2014CacheenabledSC} at each BS, or randomly cache contents in an i.i.d. manner \cite{24Bharath2016ALA} according to content popularity, or cache contents according to a uniform distribution \cite{23TamoorulHassan2015ModelingAA} at each BS. Note that the fundamental strategies introduced in \cite{22Batug2014CacheenabledSC,24Bharath2016ALA,23TamoorulHassan2015ModelingAA} may not obtain an optimal performance. Therefore,
%	, which provides no file diversity;, which may cause a waste of storage by storing multiple copies of one content at a BS;, which does not take into account the content popularity	To optimize the network performance, 	
the authors in \cite{12Cui2015AnalysisAO} consider an optimal random caching design in large-scale wireless networks and maximize the successful transmission probability (STP). However, simply considering content placement strategy design may not meet the received signal-to-interference-plus-noise ratio (SINR) requirement of the user, especially when the requested content is not stored at neighboring BSs, where a user will be associated with a relatively farther BS for serving, leading to a weak signal strength compared to the interference. As a result, some other works have introduced different delivery strategies to improve network performance.
%Note that in the aforementioned works, when the requested content is not stored at neighboring BSs, a user will be associated with a relatively farther BS for serving, leading to a weak signal strength compared to the interference. Therefore, designing a proper content delivery strategy to improve the received signal-to-interference-plus-noise ratio (SINR) also plays an important role in performance improvement.
 In \cite{015Cui2016AnalysisAO}, the authors jointly design random caching and multicasting in large-scale heterogeneous cellular networks (HCNs) and maximize the STP by optimizing cache distributions. In \cite{Xing2017TemporalSpatialRA}, a periodic discontinuous transmission scheme is proposed to improve the STP at the expense of latency. The works in \cite{018jiang2018partition} and \cite{09DongDongjiangEnMultiAntenna} are concerned with successive interference cancellation and multiple receive antennas at the user to cancel interference, respectively.	
	
	BS joint transmission (JT), as one of downlink coordinated multipoint transmission (CoMP) transmission techniques to mitigate inter-cell interference (ICI) \cite{Cui2014EvolutionOL}, is also a promising technology that can be employed to enhance the received SINR and further improve the STP in cache-enabled networks. In BS JT, requested data is shared among cooperative BSs via backhaul links, and jointly transmitted to user. According to whether the channel state information (CSI) between the cooperative BSs and the given user is available, BS JT can be divided into non-coherent joint transmission (NC-JT) \cite{003Tanbourgi2014ATM,02Nigam2014CoordinatedMJ,004Nie2016UserCentricCB} and coherent joint transmission (C-JT) \cite{007Xia2012DownlinkCM,009Kerret2015RegularizedZI,010Li2012PerformanceEO,07Yu2018CoherentJT}. In NC-JT, data is transmitted simultaneously to the user by the cooperative BSs without prior phase mismatch correction and tight synchronization \cite{003Tanbourgi2014ATM}. At the user, the sum of the desired non-coherent signals yields a received power boost to improve received SINR. Due to low complexity, NC-JT has been widely analyzed using stochastic geometry \cite{003Tanbourgi2014ATM,02Nigam2014CoordinatedMJ,004Nie2016UserCentricCB}. In \cite{003Tanbourgi2014ATM}, the authors propose a tractable model for analyzing NC-JT and depict the SINR characteristic, where the locations of BSs are modeled as Poisson point process (PPP). The authors in \cite{02Nigam2014CoordinatedMJ} introduce NC-JT into HCNs, and analyze the coverage probability of the general user and the worst user. In \cite{004Nie2016UserCentricCB}, an optimization on received signal strength thresholds with a minimum spectral efficiency constraint is performed under a cluster JT scheme. Comparing with NC-JT, C-JT can fully exploit the potential of BS JT at the expense of CSI sharing \cite{007Xia2012DownlinkCM}.	
	%However, the full potential of BS cooperation can only be exploited at the expense of sharing CSI among the cooperative BSs, i.e., C-JT \cite{007Xia2012DownlinkCM}. 	Comparing with NC-JT, 
	The main research interests of C-JT focus on the impact of imperfect CSI \cite{009Kerret2015RegularizedZI,010Li2012PerformanceEO}, and that of different amounts of CSI available among the cooperative BSs \cite{07Yu2018CoherentJT}. Specifically, \cite{009Kerret2015RegularizedZI} investigates the impact of precoding with imperfect global CSI, which is caused by CSI feedback limitation and backhaul sharing limitation. In \cite{010Li2012PerformanceEO}, the authors evaluate the performance of BS JT with predicted CSI. The effects of backhaul latency and user mobility are considered. \cite{07Yu2018CoherentJT} studies C-JT in a downlink HCN with perfect CSI and provides an approximated STP for general users. Note that the data sharing among the cooperative BSs imposes a significant pressure on backhaul networks regardless of NC-JT or C-JT.
	
	The advantages of caching in alleviating the network traffic load and those of BS JT in mitigating ICI inspire researchers to jointly design the two techniques \cite{019ao2015distributed,021Chen2016CooperativeCA,020wu2017base,11Wen2017RandomCB}. In \cite{019ao2015distributed}, the authors propose a new scheme to combine distributed caching and cooperative transmission to accelerate content delivery with C-JT under the assumption that the distances between the cooperative BSs and a typical user are identical, which can not reflect the overall performance of networks. In \cite{021Chen2016CooperativeCA}, the caching storage is divided into two portions, one of which stores the most popular contents, and the other one cooperatively stores less popular contents in different BSs. Only locally optimal caching distributions are obtained in that work without providing any insights into globally optimal design. In \cite{020wu2017base}, the user is served cooperatively based on the energy states and the cached contents at BSs, where the user capacity and coverage performance are maximized. The work just considers most popular cache strategy, which does not provide any file diversity. In \cite{11Wen2017RandomCB}, the authors propose two BS cooperative transmission policies under random caching at BSs with the content placement probability as a design parameter and maximize the STP under each scheme. Note that in \cite{021Chen2016CooperativeCA,020wu2017base,11Wen2017RandomCB}, the authors only consider the NC-JT scheme. 
	
	So far, lack of attention has been paid to the role of C-JT in mitigating ICI in cache-enabled networks. Even fewer studies investigate the relationship between NC-JT and C-JT, as well as provide a uniform design insight for both schemes in cache-enabled networks. This paper therefore sets out to figure out how BS JT and caching can jointly improve the STP, and reveal the relationship between NC-JT and C-JT in cache-enabled networks. Furthermore, an optimal probabilistic content placement strategy is obtained to provide a uniform design insight. The main contributions of this paper are summarized as follows.
	% Specifically, we consider a cooperative serving scenario in cache-enabled networks, where the probabilistic placement strategy is optimized, and NC-JT and C-JT transmission schemes are employed respectively. 
	%In this paper, we study how BS JT and caching can jointly improve the STP. The relationship between NC-JT and C-JT in cache-enabled networks is revealed. An optimal probabilistic content placement strategy is obtained to provide a uniform design insight. The main contributions of this paper are summarized as follows.
	\begin{itemize}
		\item We consider a cooperative serving scenario where files are stored at BSs using a probabilistic placement strategy, and two joint transmission schemes are involved, i.e., NC-JT and C-JT. To our best knowledge, this is the first work to employ the two schemes in cache-enabled networks to improve the STP of the typical user, respectively;
		\item We derive the STPs of the NC-JT and C-JT schemes using stochastic geometry, and reveal the relationship of the STPs between them. Specifically, for NC-JT, we present a tractable expression for the STP in the interference-limited regime. For C-JT, it is difficult to obtain a closed-form expression for the STP. Hence, we derive an upper bound and a tight approximation for the STP in this scheme, both of which have similar forms to the STP in NC-JT;
		\item We maximize the STP in NC-JT and the tight approximation of STP in C-JT by optimizing the placement probability vector. We formulate one uniform optimization problem for NC-JT and C-JT due to the similarity of the two objective functions. By exploring the properties of the STP, we propose an algorithm to obtain globally optimal solutions for several special scenarios and locally optimal solutions in the general scenario;
		\item Finally, we compare the optimized probabilistic placement strategy with three baseline strategies for NC-JT and C-JT, respectively. We show that in each transmission scheme, the optimal placement strategy achieves a significant gain in STP over the baselines. Furthermore, the STP performance in C-JT outperforms that in NC-JT for each placement strategy.
	\end{itemize}
	
	The rest of the paper is organized as follows. In Section \ref{section: Problem statement}, we describe the system model and the JT scheme considered in this paper. In Section \ref{section: STP Derivation}, we derive the expressions of the STPs in NC-JT and C-JT. In Section \ref{section: STP maximization}, we analyze and maximize the performance in both schemes. The numerical results are provided in Section \ref{section: Simulation results}, and the conclusion is drawn in Section \ref{section: Conclusion}.
	
	%另外，A1给出了考虑CSI情况下的合作增益，表明了CSI的重要性。
	%本文的contribution：不仅要突出提出的最优缓存策略的好处，也要提出CSI相比于没有CSI情况下的STP的提升。

\section{System Model and Joint Transmission Scheme}\label{section: Problem statement}
	\subsection{Network and Caching Model}
	We consider a downlink cellular network, where the BSs have limited cache storage, and several BSs cooperatively transmit the requested contents to their associated users. The BSs are randomly deployed and their locations are modeled as a 2-D homogeneous PPP $\Phi_{b}$ with density $\lambda_{b}$ \cite{02Nigam2014CoordinatedMJ}. The transmission power of each BS is denoted by $P_{b}$. According to Slivnyak's theorem \cite{01Haenggi2012StochasticGF}, the statistic observed at a randomly chosen point in a PPP $\Phi$ is the same as that observed at the origin in process $\Phi \cup \{0\}$. In this way, we focus on a typical user $u_0$, which we assume is located at the origin without loss of generality. For the wireless channel, we consider both large-scale fading and small-scale fading, and hence, the signal power received at the typical user $u_0$ from one serving BS located at $x \in \mathbb{R}^2$ can be expressed as $P_{b}\|x\|^{-\alpha} h_{x}^2 w_{x}^2 $, where $\|x\|^{-\alpha}$ and $h_{x}^2$ correspond to large-scale fading and small-scale fading, respectively; $w_x$ denotes the precoder used by serving BS located at $x$, and will be further elaborated in Section \ref{Joint Transmission Strategy}; $\alpha > 2$ denotes the path-loss exponent; the complex Gaussian distributed random variable $ h_x \stackrel{d} {\sim} \mathcal {CN}(0,1)$ or $\left|h\right|^{2}\stackrel{d}{\sim}$ Exp(1)$ $ models Rayleigh fading \cite{02Nigam2014CoordinatedMJ}. Time is divided into equal-duration time slots, and we just study one slot of the transmission.
	
	In this paper, we consider a content database containing $N \geq 1$ files, which is denoted by $\mathcal{N} = \{1,2,\cdots,N\}$. For ease of analysis, we assume that all files' sizes are equal and normalized to one, as in \cite{09DongDongjiangEnMultiAntenna, 11Wen2017RandomCB}.\footnote{Large size files can be divided into many units of equal sizes, and these units can be cached at BSs. Without loss of generality, we set the size of each unit to be one.} Each file $n\in \mathcal{N}$ has its own popularity $a_n \in [0,1]$ so that $\sum_{n \in \mathbb{N}} a_n = 1$. Here, we assume the popularity distribution follows a Zipf distribution\footnote{Note that the file popularity profile is not necessary to follow a Zipf distribution, any other proper distribution can be adopted.} \cite{11Wen2017RandomCB, 12Cui2015AnalysisAO}, i.e., $a_n = \frac{n^{-\gamma}}{\sum_{n\in \mathcal{N}} n^{-\gamma}},\, \text{for }\forall n\in \mathcal{N},$
%	\begin{equation}
%	a_n = \frac{n^{-\gamma}}{\sum_{n\in \mathcal{N}} n^{-\gamma}}, \quad \text{for }\forall n\in \mathcal{N},
%	\end{equation}
	where the parameter $\gamma \geq 0$ is the Zipf exponent, representing the popularity skewness. In this case, a file with a lower index has higher popularity, i.e., $a_1 \geq a_2 \geq \cdots \geq a_N $. The file popularity distribution $\mathbf{a} \triangleq (a_n) _{n \in \mathcal{N}} $, which can be estimated using learning methods \cite{15Golrezaei2011FemtoCachingWV}, is assumed to be known \textit{a prior} and is identical among all users. Each user randomly requests one file in a time slot according to the file popularity distribution.
	
	Each BS in the network has a limited cache size $K$ and can store $K$ different files out of $N$. It is assumed that one BS can not cache all the files in the content database, i.e., $K<N$, which is reasonable in practice. Consider a probabilistic content placement strategy, in which each BS randomly chooses $K$ files to store. Denote by $T_n \in [0,1]$ the probability that file $n$ is stored at one BS, and by $\mathbf{T} \triangleq (T_n)_{n\in\mathcal{N}} $ be the \textit{placement probability vector}, which is identical for all the BSs in the network. Then, we have \cite{12Cui2015AnalysisAO}:
	\begin{eqnarray}
	&0 \leq T_{n} \leq 1,& \label{equ: constraint of T 1}\\
	&\sum_{n \in \mathcal{N}} T_{n} = K.& \label{equ: constraint of T 2}
	\end{eqnarray}
	Based on the placement probability vector $\mathbf{T}$, each BS can randomly cache one file combination containing $K$ different files using the probabilistic content caching policy proposed in \cite{17Blaszczyszyn2015OptimalGC}. In general, the successful transmission probability of the system depends on $\mathbf{T}$, and one of our objectives is to optimizing $\mathbf{T}$ so as to maximize the STP. According to \cite{01Haenggi2012StochasticGF}, the BSs storing file $n$ can be modeled as a thinned homogeneous PPP with density $\lambda_{b} T_n$ and their locations are denoted by $\Phi_{b,n}$. Thus we have $\Phi_{b} \triangleq \bigcup_{n \in \mathcal{N}} \Phi_{b,n}$. Similarly, let $\Phi_{b,-n}$ be the set of the BSs that do not store file $n$, and thus, it is also a homogeneous PPP with density $(1-T_n)\lambda_b$. In addition, we have $ \Phi_{b,n} \bigcup \Phi_{b,-n} = \Phi_{b}$.
	\begin{figure}[!t]	
		\centering
		\includegraphics[scale=0.37]{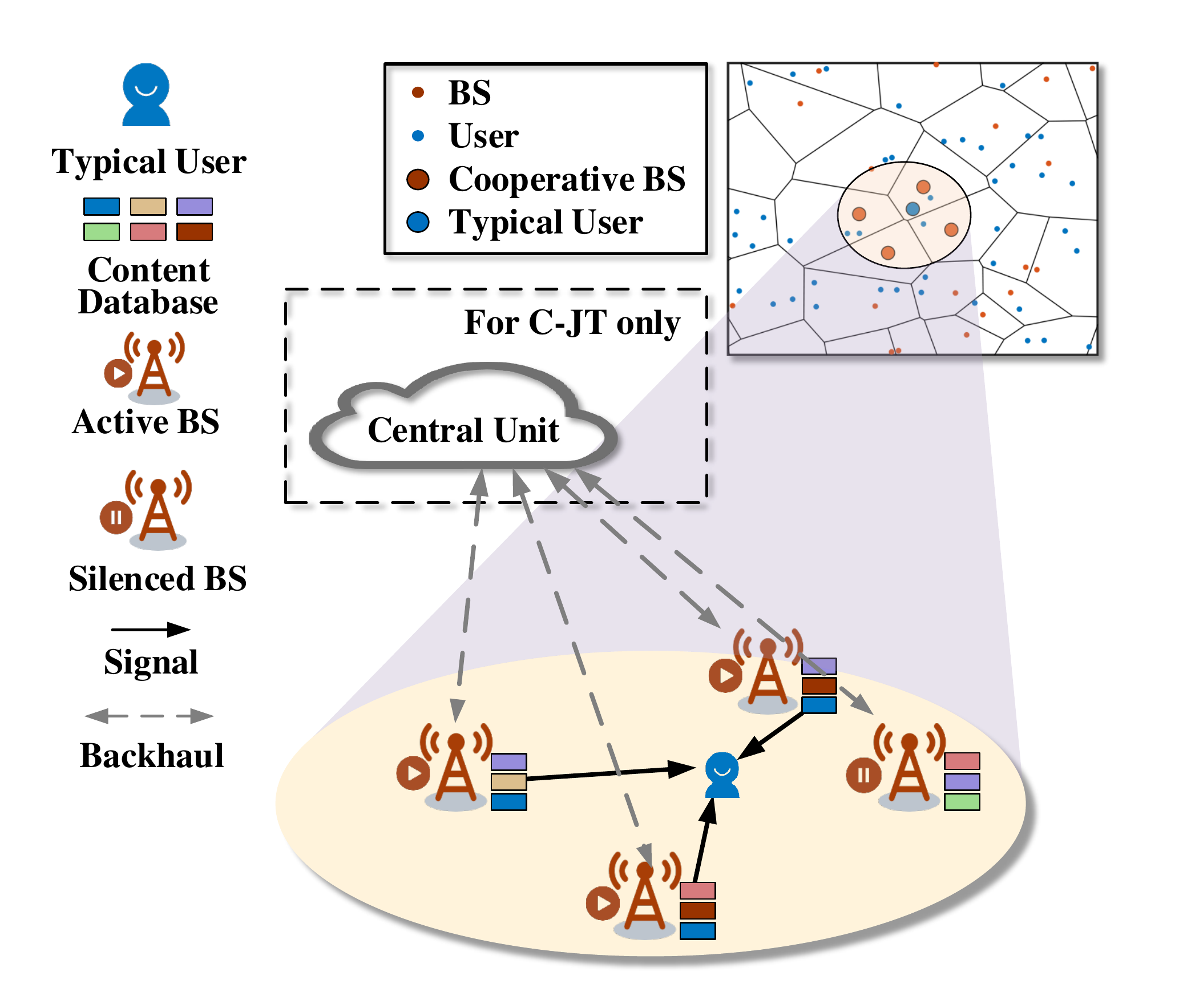} 
		\caption{Illustration of a cache-enabled cellular network with BS cooperation in NC-JT and C-JT. There are six different files ($N=6$) indicated in six different colors in the network. The color of the typical user represents the file it requests. The central unit is only used for CSI sharing in C-JT. In this scenario, $M = 4$, $K = 3$, $\left|\mathcal{C}_{n}\right| = 3$.}
		\label{fig: System Picture}
	\end{figure}
	\subsection{Joint Transmission Scheme} \label{Joint Transmission Strategy}
	It is assumed that each BS and user is equipped with a single antenna, which means only one user can be served in each frequency. Simultaneous content requests at the BSs can be handled using orthogonal multiple access methods, e.g., FDMA \cite{021Chen2016CooperativeCA}. Consider that user $u_0$ requests file $n$. We adopt a cooperative transmission policy that involves $u_0$'s $M$ nearest BSs. The set of these $M$ BSs is denoted by $\mathcal{C}$. Let $\Phi_{b}^{c}$ denote the remaining BSs that are not in the set $\mathcal{C}$, i.e., $\Phi_{b}^{c} \triangleq \Phi_{b}\backslash \mathcal{C}$, and let $\mathcal{C}_{n} \triangleq \mathcal{C} \cap \Phi_{b,n}$ denote the set of BSs that store file $n$ in $\mathcal{C}$. We refer to $\mathcal{C}_{n}$ as the \textit{cooperative set}, and denote $\mathcal{C}_{-n} \triangleq \mathcal{C} \backslash \mathcal{C}_{n}$. Note that $\left|\mathcal{C}_n\right| \leq \left|\mathcal{C}\right|=M$ and $\left|\mathcal{C}_n\right| + \left|\mathcal{C}_{-n}\right|=M$. Consider a content-centric association \cite{12Cui2015AnalysisAO}, where the serving BS of a user must store the requested file of the user but may not be its geographically nearest BS. Based on this association principle, we propose the following cooperative transmission policy:
	\begin{enumerate}
		\item If $\left|\mathcal{C}_{n}\right|=M$, it means all the $M$ BSs in $\mathcal{C}$ store the requested file $n$. In this case, all the BSs in $\mathcal{C}$ jointly transmit file $n$ to user $u_0$;\label{scheme: item1}
		
		\item If $\left|\mathcal{C}_{n}\right|\in [1,M)$, it means only part of the BSs in $\mathcal{C}$ store the requested file $n$. In this case, the BSs in set $\mathcal{C}_{n}$ jointly serve user $u_0$, and the BSs in set $\mathcal{C}_{-n}$ become silent; and\label{scheme: item2}
		
		\item If $\left|\mathcal{C}_{n}\right|= 0$, this means there are no BSs in $\mathcal{C}$ that store file $n$, then user $u_0$ will be associated with the nearest BS that stores the file, and all the BSs in $\mathcal{C}$ become silent.\label{scheme: item3}
	\end{enumerate}
	
	In all the cases above, we assume the BSs in $\Phi_{b}^{c}$ are active. Considering that they are not serving BSs of user $u_0$, we refer to them as interfering BSs. Note that in cases \ref{scheme: item2}) and \ref{scheme: item3}), the BSs in $\mathcal{C}$ that do not store the requested file $n$ are the dominant interfering BSs of $u_0$, so that silencing them can significantly reduce the interference and facilitate the transmission. Besides, there is a circumstance that the silenced BSs in $\mathcal{C}_{-n}$ of user $u_0$ may also be the serving BSs of other users, to address this problem, the coordinated scheduling method proposed in \cite{021Chen2016CooperativeCA} can be applied. In order to obtain first-order insights, we assume there is no other user served by the BSs in $\mathcal{C}_{-n}$, which can lead to an optimistic performance result. Note that it is possible that the file requested by users is not cached in any of the BSs because of the limited BS storage, which is referred to as a \textit{cache miss case}. In such a case, BSs can serve the user by retrieving the file from the core network through backhaul links, which can involve extra backhaul overhead and delay. In order to focus on the analysis of the performance of the cache-enabled network, we omit this case for simplicity and just regard it as a transmission failure.
	
	Under the scenario described above, the received signal at the typical user in a time slot can be written as
	\begin{equation}
	\begin{aligned}
	y =& \underbrace{\sum_{x \in \mathcal{C}_{n}} P_{b}^{1 / 2}\|x\|^{-\alpha / 2} h_{x} w_{x} X}_{\text{desired signal}} 
	\\ & \quad \quad \quad \quad \quad + \underbrace{\sum_{x \in \Phi_{b}^{c}} P_{b}^{1 / 2}\|x\|^{-\alpha / 2} h_{x} w_{x} X_{x}}_{\text{interference}}+Z,
	\end{aligned}	 
	\end{equation}
	where $X$ denotes the desired symbol of the typical user $u_0$, which is jointly transmitted by the BSs in the cooperative set $\mathcal{C}_n$; $X_x$ denotes the symbol sent by the BSs outside $\mathcal{C}$; $Z \stackrel{d}{\sim} \mathcal{CN}\left(0, N_{0}\right)$ is a complex Gaussian random variable, which represents the background thermal noise; $\omega_x$ denotes the precoder used by BS located at $x$. Depending on whether the CSI is available to the BS at $x$, $\omega_x$ can be set as 
	\begin{equation}
	w_{x}= \begin{cases}
	1, & \mbox{CSI unavailable;} \\ \frac{h^{*}_{x}}{\left|h_{x}\right|}, & \mbox{CSI available,}
	\end{cases}
	\end{equation}
	where $h^*_x$ denotes the complex conjugate of $h_x$. Throughout this paper, we assume that all the fading coefficients $h_x$ are i.i.d.
	In the case of \textit{NC-JT scheme}, there is no CSI at BSs, the cooperative BSs in $\mathcal{C}_n$ just jointly transmit the requested file to $u_0$ and do not need prior phase mismatch correction. As for the \textit{C-JT scheme}, we assume that each BS can obtain the CSI of the link between itself and its associated users using pilot estimation. The CSI of each link is sent to a central control unit via the backhaul channel, so that the CSI is shared among all the cooperative serving BSs, and all cooperative BSs jointly transmit the precoded signal to the user with prior phase alignment. The investigations on the impact of CSI imperfection on the network performance and improving the CSI estimation accuracy are beyond the scope of this paper and have been extensively studied in \cite{20MaddahAli2010CompletelyST} and \cite{21Yin2013ACA}, respectively. To obtain the first-order design insights of caching placement strategy, we assume perfect CSI at each BS and neglect the corresponding inaccuracy and delay caused by CSI estimation and sharing, as in \cite{09DongDongjiangEnMultiAntenna} and \cite{07Yu2018CoherentJT}.
	
	In the modern cellular network, the density of BSs is fairly high, leading to high interference. In this work, we just focus on the interference-limited regime and neglect the impact of thermal noise. The signal-to-interference ratio (SIR) of the typical user $u_0$ requesting file $n$ is given by
	\begin{equation} \label{equ: SINR_n}
	\text{SIR}_{n} =\frac{\left|\sum_{x \in \mathcal{C}_{n}} P_{b}^{1 / 2}\|x\|^{-\alpha / 2} h_{x} w_{x}\right|^{2}}{\sum_{x \in \Phi_{b}^{c}} P_{b}\|x\|^{-\alpha}\left|h_{x} w_{x}\right|^{2}},
	\end{equation}
	from which we can see that because of the silencing of the BSs in $\mathcal{C}_{-n}$, the interference is only caused by the BSs outside the set $\mathcal{C}$, and the interference from different BSs is added up directly according to their power levels.
	
	\subsection{Performance Metric}
	In this paper, we introduce the key performance evaluation metric of cache-enabled cellular networks, namely the STP. It refers to the probability that a file is transmitted successfully from a BS to its user. For a requested file $n$ of user $u_0$, if the achievable transmission data rate exceeds a target threshold $r$ $[\mathrm{bps/Hz}]$, i.e., $\log_{2}(1 + \text{SIR}_n) > r$, $u_0$ can decode the file correctly. Furthermore, the STP is defined as
	\begin{equation}\label{equ: STP defined}
	q^s(\mathbf{T}) \triangleq \operatorname{Pr}\left[\text{SIR}^s \geq \tau \right] = \sum_{n \in \mathcal{N}} a_{n} q_n^s(T_n),
	\end{equation}
	where $s \in \{\text{noCSI}, \text{CSI}\}$, indicates whether CSI is available; $\tau = 2^r-1$ denotes the SIR threshold; the second equality holds due to the total probability theorem; $q_n^s(T_n) \triangleq \operatorname{Pr} \left[ \text{SIR}^s_n \geq \tau \right] $ denotes the STP when $u_0$ requests file $n$.
	
	\newcounter{TempC}
	\setcounter{TempC}{\value{equation}}
	\setcounter{equation}{12}
	\begin{figure*}[b]
		\hrulefill
	\small{	\begin{equation} \label{equ: q_n0}
		\begin{aligned}
		q_{n,0}(T_n) = \int_0 ^\infty \!\!\int_0 ^{u_0} \exp \left( -A \left( \tau, u_0 \right) - A \left( \tau \left(\frac{u_0}{u_M}\right) ^{\frac{\alpha}{2}}, u_M \left( \frac{1}{T_n} - 1 \right) \right) \right) \frac{u_M^{M-1}}{\Gamma (M)T_n^M} \mathrm{d}u_M \mathrm{d}u_0,
		\end{aligned}
		\end{equation}		}
		\begin{equation}\label{equ: R_1}
		R_{m,1}(x,\beta) =
		\left\{
		\begin{array}{ll}{\sum\limits_{j=1}^x(-1)^{(j+1)} \binom{x}{j} \!\int_{0}^{\infty} \int\limits_{\substack{\forall t_i \in [0,1] \\ i = 1,\cdots,m}} \exp\left( - A \left( \frac{j\beta\tau} {\sum_{i=1}^{m} t_i^{-\alpha/2}}, u \right) \right) } { \frac{u^ {M-1}} {\Gamma(M)}\mathrm{d}{t_1}\cdots \mathrm{d}{t_m} \mathrm{d}u,} & {m = 1,2 \cdots, M-1,}
		\\ {0,} & { m=M, }
		\end{array}\right.
		\end{equation}
		\begin{equation} \label{equ: R_2}
		R_{m,2}(x,\beta) =
		\left\{
		\begin{array}{ll}{ \int\limits_{0}^{\infty} \exp\left( - A\left( \tau , u \right) \right) \frac{u^{M-1} }{\Gamma(M) } \mathrm{d}u, } & { m=1, }
		\\ { \sum\limits_{j=1}^x(-1)^{(j+1)} \binom{x}{j} \! \int_{0}^{\infty}\!\! \int\limits_{\substack{\forall t_i \in [0,1] \\ i = 1,\cdots,m-1}} \exp \left( -A \left(\frac{j\beta \tau}{ 1+\sum_{i=1}^{m-1} t_i^{-\alpha/2} } ,u \right) \right) \frac{u^{M-1} }{\Gamma(M) } \mathrm{d}{t_1} \cdots \mathrm{d}{t_{m-1}} \mathrm{d}{u},} & {m = 2, \cdots, M.}
		\end{array}\right.
		\end{equation}		
%		\begin{equation}\label{equ: R_1}
%			R_1(m,j,\beta)	= \int_{0}^{\infty} \int\limits_{\substack{\forall t_i \in [0,1] \\ i = 1,\cdots,m}} \exp\left( - A \left( \frac{j\beta\tau} {\sum_{i=1}^{m} t_i^{-\alpha/2}}, u \right) \right)  \frac{u^{M-1}} {\Gamma(M)}\mathrm{d}{t_1}\cdots \mathrm{d}{t_m} \mathrm{d}u,
%		\end{equation}
%	
%		\begin{equation} \label{equ: R_2}
%			R_2(m,j,\beta) = \int_{0}^{\infty} \int\limits_{\substack{\forall t_i \in [0,1] \\ i = 1,\cdots,m-1}} \exp \left( -A \left( \frac{j\beta\tau}{ 1+\sum_{i=1}^{m-1} t_i^{-\alpha/2} } ,u \right) \right) \frac{u^{M-1} }{\Gamma(M) } \mathrm{d}{t_1} \cdots \mathrm{d}{t_{m-1}} \mathrm{d}{u}.
		%\end{equation}
	\end{figure*}
	\setcounter{equation}{\value{TempC}}
\section{STP Calculation in NC-JT and C-JT}\label{section: STP Derivation}
	In this section, we first derive the expression of STP for a given placement probability vector $\mathbf{T}$ in NC-JT. Then, we derive an upper bound and a tight approximation on STP in C-JT. We verify the obtained expressions using Monte Carlo simulations in both schemes.
	
	From the cooperative transmission policy introduced in Section \ref{Joint Transmission Strategy}, $q_n^s(T_n)$ given in \eqref{equ: STP defined} can be divided into two parts. When $|\mathcal{C}_n|=0$, $u_0$ is served by its nearest BS that stores file $n$; conditioning on $|\mathcal{C}_n|=0$, the corresponding conditional STP is denoted by $q_{n,0}(T_n)$. When $|\mathcal{C}_n| = m, m=1,2 \cdots,M$, $u_0$ is served by $m$ BSs cooperatively; conditioning on $|\mathcal{C}_n|=m$, the corresponding conditional STP is denoted by $q_{c,m}^{s}$. Combining these two parts and according to the total probability theory, $q^{s} ( \mathbf{T} )$ can be written as
	\begin{equation} \label{equ: q^s 2}
		\begin{aligned}
			q^{s} \!( \mathbf{T} ) \!=\!& \sum_{n \in \mathcal{N}}\!\! a_{n} q_n^{s}(T_n) 
			\\=\!& \sum_{n \in \mathcal{N}}\!\! a_{n}\! \Bigg(\!\! \operatorname{Pr} \!\left[ \left| \mathcal{C}_{n} \right| \!=\! 0\right]\! q_{n,0}(T_n)\! +\!\! \sum_{m=1}^{M}\!\! \operatorname{Pr} \left[ \left| \mathcal{C}_{n} \right| \!= \!m\right]\! q_{c,m}^{s} \!\!\Bigg)\!.
		\end{aligned}	
	\end{equation}
%		\begin{equation} \label{equ: q^s 2}
%	\begin{aligned}
%	q^{s} ( \mathbf{T} ) =& \sum_{n \in \mathcal{N}} a_{n} q_n^{s}(T_n) 
%	\\=& \sum_{n \in \mathcal{N}} a_{n} \Bigg( \operatorname{Pr} \left[ \left| \mathcal{C}_{n} \right| = 0\right] q_{n,0}(T_n) 
%	\\ & \quad \quad\quad \quad + \sum_{m=1}^{M} \operatorname{Pr} \left[ \left| \mathcal{C}_{n} \right| =m\right] q_{c,m}^{s} \Bigg).
%	\end{aligned}	
%	\end{equation}
	Next, we give the calculation of STP in NC-JT and C-JT, respectively.
	\subsection{Calculation of STP for NC-JT} \label{STP derivation for NC-JT}
	In this part, we derive the expression of the STP $q^{\text{noCSI}} \left( \mathbf{T} \right)$ in NC-JT using stochastic geometry. In the absence of CSI, $w_x = 1$. Conditioning on having $m$ BSs in the cooperative set $\mathcal{C}_n$, the signal power received at $u_0$ is $ S^{\text{noCSI}} = \left| \sum_{i=1}^{m} \| x_i \| ^{-\alpha/2} h_{x_i} \right|^2 $, which is normalized by the BS transmission power $P_b$. For the calculation of $q_{n,0}(T_n)$ shown in \eqref{equ: q^s 2}, two types of interferers in $\Phi_{b}^{c}$ need to be considered \cite{12Cui2015AnalysisAO}: i) the BSs that store file $n$ (which are farther away from $u_0$ than the serving BSs), and ii) the BSs that do not store file $n$ (which can be closer to $u_0$ than the serving BS). To compute $q_{c,m}^{\text{noCSI}}$, we need to consider two cases: i) the $K$-th nearest BS stores file $n$ and joins the cooperative transmission, and ii) the $K$-th nearest BS does not store file $n$ and is silenced. To derive the expression of $q^{\text{noCSI}} ( \mathbf{T} )$, we first have the following lemma.
	\begin{lemma}[Joint PDF of the Distances of the Serving BSs]\label{lemma Joint pdf of BSs distances}
	 Let $R_i,i\in \mathbb{N}^+$ denote the distance between the $i$-th nearest BS and $u_0$. Then, the joint probability density function (PDF) of $\mathbf{R} = \left( R_{1}, \cdots, R_{m}, R_{M} \right) $, $m = 1,2,\cdots,M-1$; $M\in\mathbb{N}^+$ is given as
		\begin{equation}
		\begin{aligned}
		f_{\mathbf{R}}(\mathbf{r})=\frac{ 2(\pi\lambda_b)^M } {(M-1)!} r_M^{2M-1} e^{-\pi\lambda_b r_M^2 } \prod_{i=1} ^{m} \frac{2r_i} {r_M^2}, \label{equ:f_R(r) whole}
		\end{aligned}
		\end{equation}
		where $r_i\in[0,+\infty), i \in\mathbb{N}^+$ denotes the value of the corresponding variable $R_i,i\in \mathbb{N}^+$.
	\end{lemma}
	\begin{IEEEproof}
		Refer to Appendix \ref{appendix: Proof Joint pdf of BSs distance}.
	\end{IEEEproof}
	
	For ease of notation, we first define the following function:
	\begin{equation}\label{equ: A}
	A( \theta, u ) = \frac{2 u \theta}{\alpha-2}F_G(\alpha,\theta)  + u,
	\end{equation}
	where $\alpha$ is the path-loss exponent, and $F_G(\alpha,\theta) \triangleq {_2F_1}\left( 1,1-\frac{2}{\alpha}, 2-\frac{2} {\alpha} ,- \theta \right)$ denotes the Gauss hypergeometric function.
	Then, based on Lemma \ref{lemma Joint pdf of BSs distances} and using stochastic geometry, we have the following theorem.
	\begin{theorem}[STP in NC-JT] \label{theorem: STP in NC-JT}
		The STP in the case of NC-JT is given by
		\begin{equation}
		q^{\text{noCSI}} ( \mathbf{T} ) = \sum_{n \in \mathcal{N}} a_{n} q_n^{\text{noCSI}}(T_n),
		\end{equation}
		where $q_n^{\text{noCSI}}(T_n)$ is expressed as
		\begin{equation} \label{equ: qnNoCSI in theorem 1}
		\begin{aligned}
		q_n^{\text{noCSI}}(T_n) =&\underbrace{ \left( 1 - T_{n} \right) ^ {M} q_{n,0}(T_n) }_{\triangleq Q_1(T_n)}
		\\& \quad+ \underbrace{ \sum_{m=1}^{M} \dbinom{M}{m} T_{n}^{m} \left( 1-T_{n} \right) ^{M-m} q_{c,m}^{\text{noCSI}} }_{\triangleq Q^{\text{noCSI}}_2(T_n)}.
		\end{aligned}		
		\end{equation}
		Here $q_{n,0}(T_n)$ is in \eqref{equ: q_n0},	
		where $\Gamma(\cdot)$ denotes the complete Gamma function and $q_{c,m}^{\text{noCSI}} \triangleq \left(1-\frac{m}{M} \right)  R_{m,1}(1,1)+ \frac{m}{M} R_{m,2}(1,1)$ with $R_{m,1}(x,\beta)$ and $R_{m,2}(x,\beta)$ given by \eqref{equ: R_1} and \eqref{equ: R_2}, respectively.				 	
	\end{theorem}
	\begin{IEEEproof}
		Refer to Appendix \ref{appendix: Proof STP in NC-JT}.
	\end{IEEEproof}
	
	From Theorem \ref{theorem: STP in NC-JT}, we can know that $q_{n,0}(T_n)$ and $q_{c,m}^{\text{noCSI}}$ denote the STPs of $m=0$ and $m=1,2,\cdots,M$, respectively. Due to the cooperative transmission policy we adopted, $q_{n,0}(T_n)$ is a function of $T_n$ while $q_{c,m}^{\text{noCSI}}$ is not. More specifically, when $m=0$, there is no available BS in $\mathcal{C}_n$, and the nearest BS in $\Phi_{b}^{c}$ storing file $n$ needs to be found as the serving BS based on $T_n$, leading to the relevance between $q_{n,0}(T_n)$ and $T_n$. However, when $m=1,2,\cdots,M$, it is certain that there are $m$ BSs in $\mathcal{C}$ storing file $n$, therefore, $q_{c,m}^{\text{noCSI}}$ is unrelated to $T_n$, and the impact of $T_n$ is reflected in the term $\sum_{m=1}^{M} \dbinom{M}{m} T_{n}^{m} \left( 1-T_{n} \right) ^{M-m}$.
	\begin{figure}[!t]	
		\centering
		\includegraphics[height=5.5cm,width=8.8cm]{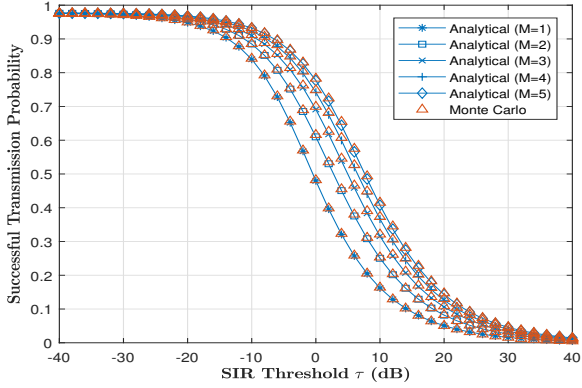} 
		\caption{STP $q^{\text{noCSI}} ( \mathbf{T} )$ versus $\tau$. $N = 8,$ $K=3,$ $\alpha = 4,$ $\lambda_b = 0.01,$ $\mathbf{T} = [0.9,0.8,0.6,0.4,0.2,0.1,0,0],$ and $\gamma=2.$ In the Monte Carlo simulations, the BSs are deployed in a square area of $1,000\times 1,000\, \mathrm{m}^2$, and the results are obtained by averaging over $10^5$ independent realizations.}
		\label{fig: SIR tau mento ana}
	\end{figure}

	Fig. \ref{fig: SIR tau mento ana} plots the STP $q^{\text{noCSI}} ( \mathbf{T} )$ versus $\tau$ with different $M$. From Fig. \ref{fig: SIR tau mento ana}, we can see that the derived analytical expression of STP in NC-JT with different $M$ matches the corresponding Monte Carlo results perfectly, which verifies Theorem \ref{theorem: STP in NC-JT}. In addition, involving one more BS to cooperatively serve $u_0$ can yield a higher STP, but the marginal gain on STP decreases with $M$. The STP $q^{\text{noCSI}} ( \mathbf{T} )$ increases as the SIR threshold $\tau$ decreases and asymptotically approaches a constant less than 1. The gap between the constant value and one is due to the cache miss case described in Section \ref{Joint Transmission Strategy}.

	\subsection{Upper Bound and Approximation on STP for C-JT} \label{upperbound and approximation for NC-JT}
		In this part, we derive an upper bound and an approximation of STP $q^{\text{CSI}} \left( \mathbf{T} \right)$ in C-JT. In the presence of CSI, $w_x = h^{*}_{x}/\left|h_{x}\right|$. Conditioning on having $m$ BSs in the cooperative set $\mathcal{C}_n$, the normalized signal power received at $u_0$ is $ S^{\text{CSI}} = \left| \sum_{i=1}^{m} \| x_i \| ^{-\alpha/2} \left|h_{x_i}\right| \right|^2 $. According to \eqref{equ: q^s 2}, $q_{n,0}(T_n)$ is given by \eqref{equ: q_n0}, and we need to derive $q_{c,m}^{\text{CSI}} $. In the process of calculating $q_{c,m}^{\text{CSI}}$, a conditional complementary cumulative density probability (CDF) $ \operatorname{Pr} \left[ S^{\text{CSI}} \geq \tau I \left| \right.\mathbf{R} = \mathbf{r},I\right] $ need to be considered. Since $\left|h_{x}\right|$ is a Rayleigh distributed random variable (RV), $S^{\text{CSI}}$ is the square of the weighted sum of Rayleigh RVs, whose CDF still can not be expressed explicitly. However, we can obtain an upper and an approximation of $q_{c,m}^{\text{CSI}}$. First, we present two lemmas.
	\begin{lemma}[A Lower Bound for the Weighted Sum of Rayleigh RVs] \label{lemma: L Bound on WSR RVs}
		Consider a sequence of i.i.d. Rayleigh RVs $X_l, l= 1,\cdots, L$ with scale parameter $\sigma$, a lower bound for the CDF of the square of their weighted sum is \cite{07Yu2018CoherentJT}
		\setcounter{equation}{15}
		\begin{equation}
		\operatorname{Pr}\left[\left(\sum_{l=1}^{L} a_{l} X_{l}\right)^{2} \leq x\right] \geq \operatorname{Pr}\left[\sum_{l=1}^{L} X_{l}^{2} \leq \frac{x}{\omega}\right],
		\end{equation}
		where $a_l \in \mathbb{R}^{+}$, $\omega \triangleq \sum_{l=1}^{L} a_l^2$, and $\sum_{l=1}^{L} X_{l}^{2} \overset{d}{\sim} {Gamma}\left(L, 2 \sigma^{2}\right)$, which denotes the gamma distribution with shape parameter $k$ and scale parameter $\theta$.
	\end{lemma}

		The CDF of a RV $X$ with distribution $Gamma(L,1)$ is $ F_{X}(x ;\! L,\! 1)\!=\!\frac{\gamma(L, x)}{\Gamma(L)}, L \!\in\! \mathbb{N}^{+} $, here $ \gamma\!\left( L, x \right) \!\triangleq \!\int_{0}^{x} \!t^{L-\!1} e^{-t} \mathrm{ d } t$ denotes the lower incomplete gamma function. Thus, $ F_{X}(x ;\! L, \!1) $ is the normalized lower incomplete gamma function.
	\begin{lemma}[Two Bounds] \label{lemma: Bound on IGF}
		The normalized lower incomplete gamma function $ F_{X}(x ; L, 1) $ is bounded as \cite{08Alzer1997OnSI}
		\begin{equation}\label{equ: BoundsonIGF}
		\left(1-e^{-\beta x}\right)^{L} \leq F_{X}(x ; L, 1) \leq \left(1-e^{-x}\right)^{L}, 
		\end{equation}
		here $\beta\! \triangleq\! \Gamma(L \!+ \!1)^{-1 / L}$. The equality holds if and only if $L\!=\!1$.
	\end{lemma}
	
		Based on Lemma \ref{lemma: L Bound on WSR RVs} and Lemma \ref{lemma: Bound on IGF}, the upper bound and approximation of $q^{\text{CSI}} ( \mathbf{T} ) $ are given by the following theorem.
	\begin{theorem}[Upper Bound and Approximation of STP in C-JT]\label{theorem: upper and Appr of C-JT} 
		The upper bound of the STP $q^{\text{CSI}} ( \mathbf{T} ) $ in C-JT is given by
		\begin{equation}\label{equ: qnCSIupper in theorem 3}
		q^{\text{CSI},u} ( \mathbf{T} ) = \sum_{n \in \mathcal{N}} a_{n} q_n^{\text{CSI},u}(T_n),
		\end{equation}
		where $q_n^{\text{CSI},u}(T_n)$ is given by
		\begin{equation}\label{equ: qnCSIupper2 in theorem 3}
		q_n^{\text{CSI},u}(T_n) \!=\! Q_1(T_n)\! +\! \sum_{m=1}^{M}\! \dbinom{M}{m}\! T_{n}^{m} \!\left( 1\!-\!T_{n} \right) ^{M\!-\!m} \!q_{c,m}^{\text{CSI},u} .	
		\end{equation}
		Here, $Q_1(T_n)$ is given by \eqref{equ: qnNoCSI in theorem 1}, and $q_{c,m}^{\text{CSI},u} \triangleq \left(1-\frac{m}{M} \right) R_{m,1}(m,\beta) + \frac{m}{M} R_{m,2}(m,\beta)$, with $\beta = \Gamma(m+1)^{-1 / m}$. $R_{m,1}(x,\beta)$ and $R_{m,2}(x,\beta)$ are given by \eqref{equ: R_1} and \eqref{equ: R_2}, respectively.			
		
		The approximation of $q^{\text{CSI}} ( \mathbf{T} ) $ can be easily obtained by substituting $\beta = 1$ into $q^{\text{CSI},u} ( \mathbf{T} )$, i.e.,
		\begin{equation}\label{equ: Appr of qCSI}
		q ^{\text{CSI},a} (\mathbf{T})=\left. q ^{\text{CSI},u} (\mathbf{T}) \right| _{\beta = 1},
		\end{equation}
		and the superscript ``$\text{CSI},u$'' in \eqref{equ: qnCSIupper in theorem 3} and \eqref{equ: qnCSIupper2 in theorem 3} are replaced by ``$\text{CSI},a$'' correspondingly.
	\end{theorem}
	\begin{IEEEproof}
		Refer to Appendix \ref{appendix: Proof of STP in C-JT}.
	\end{IEEEproof}	
		\begin{figure}[!t]	
		\centering
		\includegraphics[height=5.5cm,width=8.8cm]{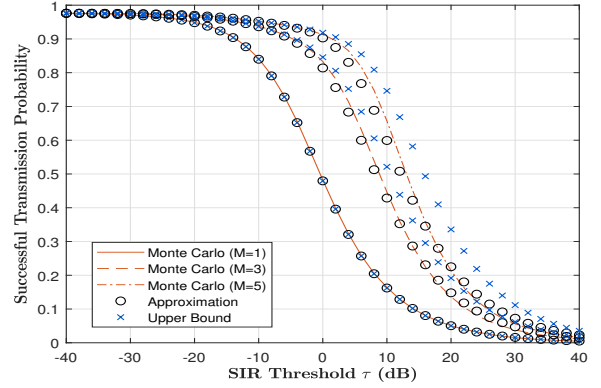} 
		\caption{The upper bounds, the approximations and the Monte Carlo results of STP in C-JT versus $\tau$. Simulations parameters are the same as in Fig. \ref{fig: SIR tau mento ana}.}
		\label{fig: UpperAppr of STP CSI}
	\end{figure}
	\begin{remark}[Relationships between the STP in NC-JT and C-JT]
		From Theorem \ref{theorem: STP in NC-JT} and Theorem \ref{theorem: upper and Appr of C-JT}, we can easily find the relationships below:
		\begin{enumerate}
			\item  $ q^{\text{CSI},u} ( \mathbf{T} ) $ and $ q^{\text{CSI},a} ( \mathbf{T} ) $ have similar forms with $ q^{\text{noCSI}} ( \mathbf{T} ) $. To obtain $ q^{\text{CSI},u} ( \mathbf{T} ) $ (or $ q^{\text{CSI},a} ( \mathbf{T} ) $), we just take an upper bound on (or make an approximation of) the STP when $m=1,2,\cdots,M$. In addition, $ q^{\text{noCSI}} ( \mathbf{T} ) $, $ q^{\text{CSI},u} ( \mathbf{T} ) $ and $ q^{\text{CSI},a} ( \mathbf{T} ) $ coincide when $M=1$, which means NC-JT and C-JT are identical and the upper bound and the approximation are tight at $M=1$. For the case of $m=0$, we do not make any operation on $Q_1(T_n)$ since it can be expressed analytically.
			\item In the case of $m=1,2,\cdots,M$, $q_{c,m}^{\text{noCSI}}$ is a special case of $q_{c,m}^{\text{CSI},u}$ and $q_{c,m}^{\text{CSI},a}$ when $x=1$ and $\beta=1$. Furthermore, $q_{c,m}^{\text{noCSI}}$, $q_{c,m}^{\text{CSI},u}$ and $q_{c,m}^{\text{CSI},a}$ are unrelated to $T_n$, and the impact of $T_n$ is reflected in the term {\small$\sum_{m=1}^{M} \dbinom{M}{m}T_{n}^{m} \left( 1-T_{n} \right) ^{M-m}$}. Therefore, when select $T_n,n \in \mathcal{N}$ as the optimization variables, the coefficients $q_{c,m}^{\text{noCSI}}$, $q_{c,m}^{\text{CSI},u}$ and $q_{c,m}^{\text{CSI},a}$ will not change the properties of the STP. As a result, the optimization problems in both schemes can be formulated uniformly, which will be shown in Section \ref{section: STP maximization}.
		\end{enumerate}	
	\end{remark}
	  
	  Fig. \ref{fig: UpperAppr of STP CSI} plots $ q^{\text{CSI},u} ( \mathbf{T} ) $ and $ q^{\text{CSI},a} ( \mathbf{T} ) $ versus $\tau$ at different $M$ and their corresponding Monte Carlo results. As can be seen from Fig. \ref{fig: UpperAppr of STP CSI}, $ q^{\text{CSI},u} ( \mathbf{T} ) $, $ q^{\text{CSI},a} ( \mathbf{T} ) $ and their Monte Carlo results coincide when $M=1$; when $M\geq2$, $ q^{\text{CSI},u} ( \mathbf{T} ) $ bounds the Monte Carlo results from above, and $ q^{\text{CSI},a} ( \mathbf{T} ) $ tightly approximates the Monte Carlo results over the whole range of $\tau$.		
\section{STP Maximization in NC-JT and C-JT}\label{section: STP maximization}
	In this section, we maximize the STP in NC-JT and the approximation of STP in C-JT by optimizing the placement probability vector $\mathbf{T}$. For NC-JT, we take $ q^{\text{noCSI}} ( \mathbf{T} )$ as the objective function. For C-JT, $ q^{\text{CSI}} ( \mathbf{T} ) $ can not be expressed in a tractable form. Thus, we maximize $ q^{\text{CSI},a} ( \mathbf{T} ) $, which provides a good approximation for $ q^{\text{CSI}} ( \mathbf{T} ) $ as shown in Fig. \ref{fig: UpperAppr of STP CSI}. To facilitate problem solving, we first present some properties of $ q^{\text{noCSI}} ( \mathbf{T} ) $.
	\begin{remark}[Properties of  $ q^{\text{noCSI}} ( \mathbf{T} )$] \label{remark: Properties of STP for NC-JT}
		Several properties can be easily found from the expressions of  $ q^{\text{noCSI}} ( \mathbf{T} )$:
		\begin{enumerate}
			\item The STP $q^{\text{noCSI}} ( \mathbf{T} )$ is an increasing function of $m$, i.e., $q_{c,m}^{\text{noCSI}}>q_{c,m-1}^{\text{noCSI}}$ for $m = 2,\cdots,M$. That is, involving more BSs to cooperatively transmit the same file $n$ to $u_0$ helps improve the STP;	\label{property 1}		
			\item $q_{n,0}(T_n)$ is an increasing function of $T_n$, i.e., $q^{\prime}_{n,0}(T_n)>0$, which comes from the fact that a file with a higher probability being stored at a BS has a higher STP. Furthermore, we have $q_{n,0}(1)<q_{c,1}^{\text{noCSI}}$, and hence, we have $q_{n,0}(T_n)<q_{c,1}^{\text{noCSI}}$ for all $T_n \in [0,1]$, which comes from the fact that: when $m=0$, all the $M$ BSs in $\mathcal{C}$ are silenced and the only serving BS is outside $\mathcal{C}$; however, when $m=1$, all the BSs in $\mathcal{C}$ except the only one serving BS are silenced and the serving BS is inside $\mathcal{C}$. Therefore, the distance between the serving BS and $u_0$ when $m=0$ is larger than that when $m=1$, yielding a lower STP;	\label{property 2}		
			\item When $ q_{c,m+1}^{\text{noCSI}} - q_{c,m}^{\text{noCSI}} \leq q_{c,m}^{\text{noCSI}} - q_{c,m-1}^{\text{noCSI}} $ for $m = 2,\cdots,M-1$, combining that $q_{c,m}^{\text{noCSI}} > q_{c,m-1}^{\text{noCSI}}$, we have $Q^{\text{noCSI}}_2(T_n)$ is a concave function of $T_n$ \cite{019ao2015distributed}. Besides, we have $\frac{\mathrm{ d }^2 Q_1(T_n)}{\mathrm{ d } T_n^2}<0$, which indicates $Q_1(T_n)$ is also a concave function of $T_n$, and thus, $q^{\text{noCSI}} ( \mathbf{T} )$ is concave. \label{property 3}		
		\end{enumerate}
	\end{remark}	

	Note that $ q^{\text{CSI},a} ( \mathbf{T} ) $ also possesses the above properties, due to the similarity in the form with $ q^{\text{noCSI}} ( \mathbf{T} )$.
	For ease of notation, we use a uniform variable $q^{g} ( \mathbf{T} )$ to denote the objective functions. $g$ can be set as ``noCSI'' or ``$\text{CSI},a$'', representing $ q^{\text{CSI}} ( \mathbf{T} ) $ and $ q^{\text{CSI},a} ( \mathbf{T} ) $. Then, the optimization problem can be uniformly formulated as:
	\begin{problem}[Maximization of STP $q^{g} ( \mathbf{T} )$]\label{problem: 1}
		\begin{align*}
		&\underset{\mathbf{T}}{\max} \,\, q^{g} ( \mathbf{T} )
		\\& s.t. \,\, \eqref{equ: constraint of T 1}, \eqref{equ: constraint of T 2}.
		\end{align*}		
	\end{problem}
	Let $\mathbf{T}^\star$ be the optimal solution and $q^{g\star}= q^{g}(\mathbf{\mathbf{T}^\star})$ be the optimal value.	Because of the complicated expression of $q^{g} ( \mathbf{T} )$, the convexity of the objective function in Problem \ref{problem: 1} is hard to determine. However, based on Remark \ref{remark: Properties of STP for NC-JT}, we can simplify the optimization problem significantly. When $ q_{c,m+1}^{g} - q_{c,m}^{g} \leq q_{c,m}^{g} - q_{c,m-1}^{g} $ for $m = 2,\cdots,M-1$, $q^{g} ( \mathbf{T} )$ is concave, and hence, Problem \ref{problem: 1} is a convex optimization problem, so that Slater's condition is satisfied, and we can use KKT conditions to solve this problem. Otherwise, Problem \ref{problem: 1} is a non-convex problem where the objective function is non-convex but differentiable, and the constraint set is convex. A locally optimal solution can be solved using interior point method.
	\begin{theorem}[Optimal Solution to Problem \ref{problem: 1}] \label{theorem: Solution of P1}
		The optimal solution $\mathbf{T}^\star$ to Problem \ref{problem: 1} under the condition that $ q_{c,m+1}^{g} - q_{c,m}^{g} \leq q_{c,m}^{g} - q_{c,m-1}^{g} $ for $m = 2,\cdots,M-1$ is
		\begin{equation} \label{equ: Optimal Solution of P1}
		T_{n}^{\star}=\left\{\begin{array}{ll}{0,} & {a_n D^g_n(0) < \nu,} \\ {1,} & {a_n D^g_n(1) > \nu,} \\ {x\left(T_{n}^{\star}, a_{n}, \nu \right),} & {\text { otherwise, }}\end{array}\right.
		\end{equation}
		where $D^g_n(x) \triangleq \left. \frac{\mathrm{d} q_n^{g}(T_n)}{\mathrm{d} T_n} \right| _{T_n = x}$ denotes the first-order derivative of $q_n^{g}(T_n)$ with respect to $T_n$; $x\left(T_{n}^\star, a_{n}, \nu \right)$ denotes the solution of equation $a_n D^g_n(T_n^\star) = \nu$, and $\nu$ satisfies $\sum_{n \in \mathcal{N}} T_{n}^{\star}(\nu)=K$.
	\end{theorem}
	\begin{IEEEproof}
		Refer to Appendix \ref{appendix: Proof Optimal solution of P1}.
	\end{IEEEproof}
	%qn对Tn的导数表达式==============================
	% \begin{equation} \label{equ: diff of qn}
	% \begin{aligned}
	% \frac{\mathrm{d} q_n(T_n)}{\mathrm{d} T_n}
	% & = -M(1-T_n) ^{(M-1)} q_{n,0}(T_n) + (1-T_n)^M \frac{\partial q_{n,0}(T_n)} {\partial T_n}
	% \\& +\sum_{m=1}^{M-1} \dbinom{M}{m} T_{n}^{m-1} \left( 1-T_{n} \right) ^{M-m-1} \left( m - M T_n \right) q_{c,m} + M T_n^{M-1} q_{c,M}
	% \end{aligned}
	% \end{equation}
	% and $\frac{\partial q_{n,0}(T_n)} {\partial T_n} $ is given as follow
	% \begin{equation} \label{equ: diff of qn}
	% \begin{aligned}
	% \frac{\partial q_{n,0}(T_n)} {\partial T_n}
	% & = \frac{q_{n,0} (T_n)} {T_n} + \int_0 ^\infty \int_0 ^{r_0} \frac{4 \pi ^{ M+1 } \lambda_{b} ^{ M+1 } T_{n} r_0 r_M ^{2 M-1} } { (M-1) !} \exp\left( A_1 T_n \right) \exp\left( A_2 \left( 1 - T_n \right) \right)
	% \\& \times \exp{ \left( - \tau r_0^{\alpha}\frac{N_0}{P_b}\right) } \exp \left(A_3 T_{n} + A_4 \left( 1-T_{n} \right) \right) \left( A_1 - A_2 + A_3 - A_4 \right) \mathrm{d}r_M \mathrm{d}r_0
	% \end{aligned}
	% \end{equation}
	%here $A_1 = -2 \pi \lambda_{b} \frac{r_0^2}{\alpha -2} \tau {_2F_1}\left( 1,1-\frac{2}{\alpha}; 2-\frac{2}{\alpha}, -\tau \right)$, $A_2 = -2 \pi \lambda_{b} \frac{r_M^2}{\alpha -2} \frac{s}{r _M ^\alpha} {_2F_1}\left( 1, 1-\frac{2} {\alpha}; 2-\frac{2}{\alpha}, -\frac{s}{r _M ^\alpha} \right) $, $A_3 = -\pi \lambda_{b} r_0^{2}$, and $A_4 = - \pi \lambda_{b} r_M^{2}$.
	%======================================================

	When calculating the optimal solution according to Theorem \ref{theorem: Solution of P1}, $x\left(T_{n}^\star, a_{n}, \nu \right)$ can be obtained using bisection search, since $D^g_n(x)$ is a decreasing function. Similarly, we can also obtain $\nu$ by solving the equation $\sum_{n \in \mathcal{N}} T_{n}^{\star}(\nu)=K$ via the bisection search. The corresponding procedure is summarized in Algorithm \ref{Algo: Optimal Solution of Problem 1}.
	\begin{figure}[!t]	
		\centering
		\includegraphics[height=5.5cm,width=8.8cm]{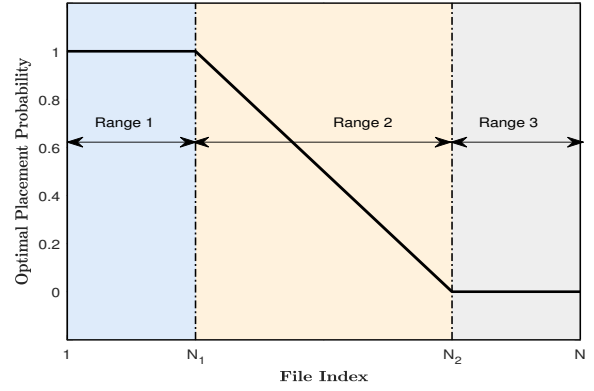} 
		\caption{The relationship between the file popularity and the optimal placement probability. Here, a file with a smaller index has higher popularity since the popularity distribution follows Zipf distribution.}
		\label{fig: T versus N}
	\end{figure}	
		\begin{algorithm}
		\caption{Optimal Solution to Problem \ref{problem: 1}} 
		\label{Algo: Optimal Solution of Problem 1}
		\begin{algorithmic}[1] 
			\IF{ $ q_{c,m+1}^{g} - q_{c,m}^{g} \leq q_{c,m}^{g} - q_{c,m-1}^{g} $ for $m = 2,\cdots,M-1$} 
			\STATE Initialize $\left[ \nu^{(0,min)}\!,\! \nu^{(0,max)} \right] = \left[ a_N D^g_n(1) , a_1 D^g_n(0) \right] $ and $\epsilon\! = \!10^{-6}$.
			\REPEAT
			\STATE $\nu^{(l+1)} \leftarrow \nu^{(l,min)} + \frac{\nu^{(l,max)} - \nu^{(l,min)}}{2}$.
			\STATE Obtain $T_n$ for all $n \in \mathcal{N}$ according to Theorem \ref{theorem: Solution of P1}.
			\IF {{\scriptsize $\left(\!\sum_{n \in \mathcal{N}}\! T_{n}\!\left(\! \nu^{(l+1)}\! \right)\!-\!K \right) \left(\sum_{n \in \mathcal{N}} \!T_{n}\!\left(\! \nu^{(l,max)}\! \right)\! -\!K \right) \!> \!0$}}
			\STATE $\nu^{(l+1,max)} \leftarrow \nu^{(l+1)}$
			\ELSE
			\STATE $\nu^{(l+1,min)} \leftarrow \nu^{(l+1)}$
			\ENDIF
			\STATE $l\leftarrow l+1.$
			\UNTIL $\nu^{(l,max)} - \nu^{(l,min)} < \epsilon$
			\ELSE
			\STATE Get $T_n$ for all $n \in \mathcal{N}$ using the interior point method.
			\ENDIF 
		\end{algorithmic} 
	\end{algorithm}
	\begin{figure}[!t]
	\centering
	\subfloat[ $\tau = 0\, \mathrm{ dB }$]{\includegraphics[height=5.5cm,width=8cm]{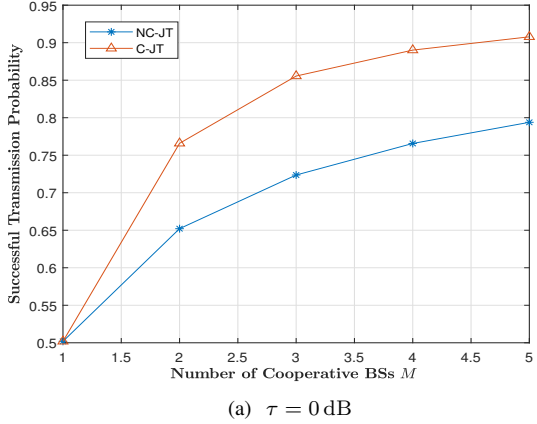} {\label{fig: Comparison NC C sub 1}}}\\	
	\subfloat[$M=3$]{\includegraphics[height=5.5cm,width=8cm]{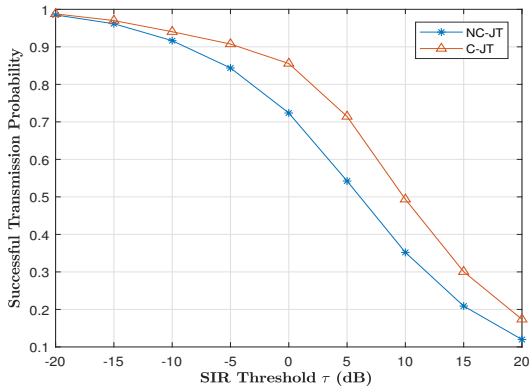}\label{fig: Comparison NC C sub 2}}
	\caption{STP Comparisons between NC-JT and C-JT. $N=8$, $K=3$, $\gamma = 2$.}
	\label{fig: Comparison NC C}
\end{figure} 
	\begin{remark}[Interpretations of Theorem \ref{theorem: Solution of P1}]\label{remark: interpretation of theorem 2}
		From Theorem \ref{theorem: Solution of P1}, we can have the following observations when the optimal solution is reached:
		\begin{enumerate}
			\item A certain $\nu$ can be obtained, which is identical to all the files. According to the equation $a_n D^g_n(T_n^\star) = \nu$, we know that the optimal solution $T_n^\star$ for a file depends on its popularity $a_n$. Since $D^g_n(x)$ is a decreasing function, a higher popularity $a_n$ will lead to a higher $T_n^\star$;
			\item There exists a scenario, in which we have $1\leq N_1<N_2\leq N$ that satisfy $T_1^\star=T_2^\star=\cdots = T_{N_1}^\star=1$, $T_{i}^\star\in (0,1)$ for all $i\in(N_1,N_2)$ and $T_{N_2}^\star=T_{N_2+1}^\star=\cdots=T_{N}^\star=0$ as illustrated in Fig. \ref{fig: T versus N}. These three cases correspond to Range 1, Range 2 and Range 3, respectively. All the files in Range 1 are cached at every BS in order to maximize the STP thanks to their high popularity, which satisfies $a_nD^g_n(1)>\nu $. By contrast, in Range 3, the files are highly unpopular and do not need to be cached at any BS. The popularity of the files in this range satisfies $a_nD^g_n(0)<\nu$. Lastly, in Range 2, the files are randomly cached at each BS to obtain more file diversity gains. 
		\end{enumerate}		
	\end{remark}

	The following lemma follows from Remark \ref{remark: interpretation of theorem 2}:
	\begin{lemma}[Property of Optimal Solution to Problem \ref{problem: 1}] \label{lemma: Property of Optimal Solution for Problem 1 }
		The optimal solution $\mathbf{T}^\star$ to Problem \ref{problem: 1} obtained by Algorithm \ref{Algo: Optimal Solution of Problem 1} satisfies $1\geq T_{1}^{\star} \geq \cdots \geq T_{N}^{\star} \geq 0$.
	\end{lemma}
	\begin{IEEEproof}
		Refer to Appendix \ref{appendix: Proof property of optimal solution for P1}.
	\end{IEEEproof}

	Lemma \ref{lemma: Property of Optimal Solution for Problem 1 } shows that a file of higher popularity can get more storage resources, which in turns helps improve the STP.
\section{Numerical results}\label{section: Simulation results}
	In this section, we conduct simulations to validate the optimality of the proposed probabilistic content placement strategy. We first compare the performances of NC-JT and C-JT based upon the optimal placement strategy. Then, we compare these two optimal designs with three baseline strategies, i.e., MPC (most popular caching) \cite{22Batug2014CacheenabledSC}, IIDC (i.i.d. caching) \cite{24Bharath2016ALA} and UDC (uniform distribution caching) \cite{23TamoorulHassan2015ModelingAA}. Note that the three baselines also adopt NC-JT and C-JT transmission schemes. We set $\alpha = 4$ and $\lambda_{b} = 0.01$. In the simulations, we obtain the optimal placement probability vector $\mathbf{T}^\star$ by using Algorithm \ref{Algo: Optimal Solution of Problem 1} for NC-JT and C-JT. Then, the optimal graphical method proposed in \cite{17Blaszczyszyn2015OptimalGC} is used to obtain the corresponding file combinations in every BS.
	\begin{figure}[!t]	
	\centering
	\includegraphics[height=5.5cm,width=8.8cm]{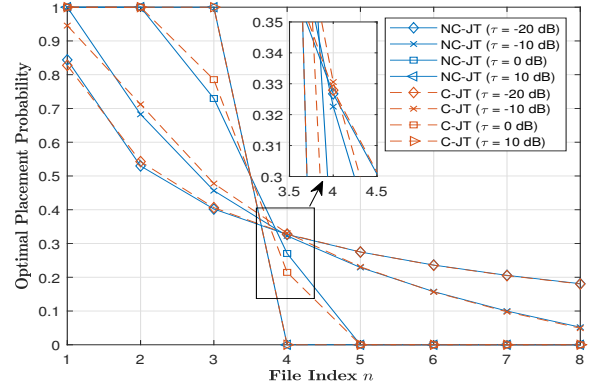} 
	\caption{Optimal placement probability $T_{n}^{\star}$ versus file index $n$ at different SIR thresholds $\tau$. $M=3$, $N=8$, $K=3$, $\gamma = 2$.}
	\label{fig: Tn-n tau changes}
	\end{figure}
	\begin{figure*}[!t]
		\centering
		\subfloat[ Number of cooperative BSs $M$ at $N=100$, $K=25$, $\tau = 0\, \mathrm{ dB }$, and $\gamma = 0.8$. ] {\includegraphics[height=4.0cm,width=5.8cm]{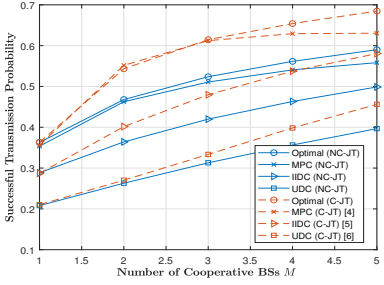}\label{fig: sub STP-M Baseline}}\,\,
		\subfloat[SIR threshod $\tau$ at $M=3$, $N=100$, $K=25$, and $\gamma = 0.8$.] {\includegraphics[height=4.0cm,width=5.8cm]{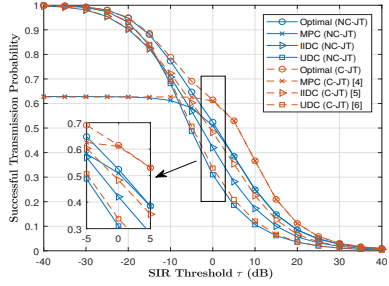}\label{fig: sub STP-tau Baseline}}\,\,
		\subfloat[Zipf exponent $\gamma$ at $M\!=\!3$, $N\!=\!100$, $K\!=\!25$, and $\tau = 0\, \mathrm{ dB }$.] {\includegraphics[height=4.0cm,width=5.8cm]{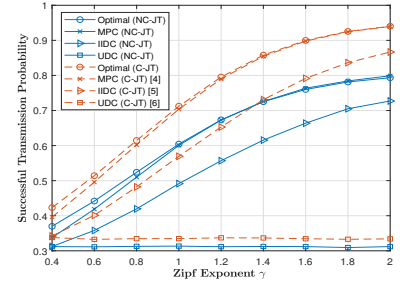}\label{fig: sub STP-gamma Baseline}}\\	
		\subfloat[Cache size $K$ at $M=3$, $N=100$, $\tau = 0\, \mathrm{ dB }$, and $\gamma = 0.8$.] {\includegraphics[height=4.0cm,width=6cm]{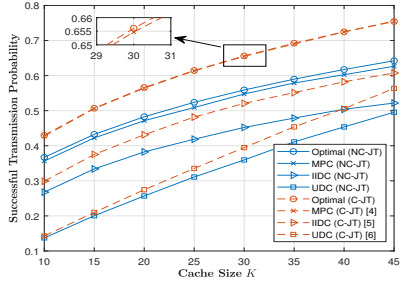}\label{fig: sub STP-K Baseline}}
		\quad
		\subfloat[Number of files $N$ at $M\!=\!3$, $K\!=\!25$, $\tau\! =\! 0\, \mathrm{ dB }$, and $\gamma = 0.8$.]{\includegraphics[height=4.0cm,width=6cm]{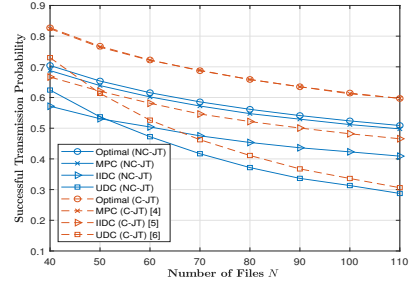}\label{fig: sub STP-N Baseline}}	
		\caption{Comparisons between proposed strategies and baselines with different number of cooperative BSs $M$, SIR threshold $\tau$, Zipf exponent $\gamma$, cache size $K$ and number of files $N$ in NC-JT and C-JT.}
		\label{fig: STP Baseline}
	\end{figure*}
	\subsection{Performance of the Proposed Probabilistic Content Placement Strategies in NC-JT and C-JT}
	In this part, we compare the performances of NC-JT and C-JT based upon the optimal probabilistic content placement strategy. Fig. \ref{fig: Comparison NC C} compares the STPs of NC-JT and C-JT versus the number of cooperative BSs $M$ and the SIR threshold $\tau$, while Fig. \ref{fig: Tn-n tau changes} depicts the corresponding $\mathbf{\mathbf{T}^\star}$ obtained by Algorithm \ref{Algo: Optimal Solution of Problem 1}, respectively. Fig. \ref{fig: Comparison NC C}\subref{fig: Comparison NC C sub 1} shows that the STPs increase with $K$ in both schemes. This is because with more BSs joining the cooperative transmission, user $u_0$ can receive a higher desire signal power and the strength of interference decreases because of the silencing of BSs in $\mathcal{C}_{-n}$. When $M=1$, the STPs in both cases are equal; when $M>1$, C-JT outperforms NC-JT, which means that the knowledge of the CSI can facilitate the transmission of files. As can be observed from Fig. \ref{fig: Comparison NC C}\subref{fig: Comparison NC C sub 2}, the STP under each scheme decreases with $\tau$. The STP gap between these two cases decreases to 0 when $\tau\rightarrow 0$, since in this regime the inequality $\text{SIR}\geq \tau$ is easy to satisfy, and the role of CSI in improving STP is no longer important.
	
	As can be seen from Fig. \ref{fig: Tn-n tau changes}, files of higher popularity tend to be stored at BSs. When $\tau$ is lower, $T_n^\star$ decreases more slowly, which is due to the fact that when $\tau$ decreases, the inequality $\text{SIR}\geq \tau$ gets easier to satisfy, therefore, storing more distinct files can increase the file diversity and further improve the STP. Besides, we can also find that when $\tau$ is large, $1\leq N_1<N_2\leq N$ holds such that $T_i^\star=1$ for $i\in[1,N_1]$ and $T_i^\star=0$, for $i\in[N_2,N]$ as illustrated in Fig. \ref{fig: T versus N}. In this situation, the proposed probabilistic content placement strategy degenerates to MPC. Besides, there is not much difference between the optimal solutions obtained for NC-JT and C-JT.
	\subsection{Comparisons Between Proposed Probabilistic Content Placement Strategy With Baselines}
	Fig. \ref{fig: STP Baseline} plots the STP versus the number of cooperative BSs $M$, SIR threshold $\tau$, Zipf exponent $\gamma$, cache size $K$ and the number of files $N$. We can see that all the STPs with the C-JT scheme outperform those with the NC-JT scheme, which confirms the effect on improving the STP by knowing the CSI. In addition, the proposed optimal placement strategy outperforms all the three baselines when adopting the same transmission scheme. More specifically, Fig. \ref{fig: STP Baseline}\subref{fig: sub STP-M Baseline} shows that the STPs of all the placement strategies increase with $M$, thanks to joint transmission and BS silencing. As can be seen from Fig. \ref{fig: STP Baseline}\subref{fig: sub STP-tau Baseline}, the STPs of all the placement strategies decrease with $\tau$. When $\tau$ is small (e.g., $\tau<-10\,\,\mathrm{ dB }$), the STP of MPC is lower than those of the other strategies. Furthermore, when $\tau\rightarrow 0$ the STPs of all the strategies except MPC asymptotically approach one, whereas the STP of MPC does not (due to MPC providing no file diversity gains). When $\tau$ is large (e.g., $\tau>0\,\,\mathrm{ dB }$), the STP of MPC, as well as that of the optimal strategy, is the highest. This is because when $\tau$ is small, storing more different files can improve the STP. However, when $\tau$ is large, ensuring the successful transmission of the most popular files makes more sense in terms of improving the STP. In addition, when $\tau>0\,\,\mathrm{ dB }$, the STP of the optimal strategy is the same as that of MPC in C-JT, and the optimal strategy degenerates to MPC as shown in Fig. \ref{fig: T versus N} and Fig. \ref{fig: Tn-n tau changes}. One can observe from Fig. \ref{fig: STP Baseline}\subref{fig: sub STP-gamma Baseline} that the STPs of all the placement strategies except UDC increase with $\gamma$, which is because that with the increase of $\gamma$, a file with a smaller index is more popular, and thus the possibility that a file requested by a user being stored at a nearby BS is higher when adopting the popularity-based placement strategy. In addition, the STP of the optimal strategy is approximately equal to that of MPC in both NC-JT and C-JT, which is because we set the parameter $\tau = 0\,\, \mathrm{ dB }$. Fig. \ref{fig: STP Baseline}\subref{fig: sub STP-K Baseline}, shows that the STPs of all the placement strategies increase with $K$, which is because increasing $K$ leads to a higher probability for a requested file to be stored at the cooperative BSs. As can be seen from Fig. \ref{fig: STP Baseline}\subref{fig: sub STP-N Baseline}, the STPs of all the placement strategies decrease with $N$, which is because increasing $N$ leads to a lower possibility for a requested file to be stored at the cooperative BSs. 
	
	\section{Conclusion}\label{section: Conclusion}
	In this paper, we studied the optimal probabilistic content placement strategy at BSs in cache-enabled cellular networks with BSs cooperation. We considered two joint transmission schemes, i.e., NC-JT and C-JT. First, we derived a tractable expression for the STP in NC-JT, and gave an upper bound and a tight approximation for the STP in C-JT, which share the similar forms. Then, we formulated a uniform optimization problem for the maximization of the STP in NC-JT and the approximation of the STP in C-JT by optimizing the placement probability vector. For both schemes, an algorithm was proposed to obtain a locally optimal solution in general cases and globally optimal solutions in several special cases. Finally, we compared the optimized performance with that of some exiting placement strategies, e.g., MPC, IIDC, and UDC. Simulation results demonstrated that the optimized placement strategy is capable of achieving a better STP performance.
	
	Although our paper assumes perfect CSI in the case of C-JT, studying the impact of CSI estimation errors and a capacity-limited backhaul channel for the sharing of CSI would be an interesting future research topic. In addition, our work can be further extended to more complicated network architectures such as HCNs.
	
	\appendices
	\section{Proof of Lemma \ref{lemma Joint pdf of BSs distances}}\label{appendix: Proof Joint pdf of BSs distance}
	We prove this lemma using the conditional probability formula. Conditioning on $R_M = r_M$, we can calculate the joint PDF with $f_{\mathbf{R}}(\mathbf{r})=f_{\left.R_1,R_2,\cdots, R_m \right| R_M} (r_1,r_2 \cdots r_m)f_{R_M}(r_M)$.
	%\begin{equation}\label{equ: appendix PDF of R}
	%f_{\mathbf{R}}(\mathbf{r})=f_{\left.R_1,R_2,\cdots, R_m \right| R_M} (r_1,r_2 \cdots r_m)f_{R_M}(r_M).
	%\end{equation}
	Here, $f_{\left.R_1,R_2,\cdots, R_m \right| R_M} (r_1,r_2 \cdots r_m) $ is the conditional joint PDF and $f_{R_M}(r_M)$ denotes the PDF of the distance between the $M$-th nearest BS and $u_0$. Since the distribution of BSs is modeled as a homogeneous PPP $\Phi_{b}$ with density $\lambda_{b}$, the PDF $f_{R_M}(r_M)$ can be given as \cite{01Haenggi2012StochasticGF}
	\begin{equation}\label{equ: appendix PDF of R_M}
	f_{R_M\!\!}\left(r_M\right)\! =\! \frac{2\left(\pi \lambda_{b}\right)^{\!M}}{\Gamma(M)} r_{M}^{2 M\!-\!1}\! e^{-\pi \lambda_{b} r_{M}^{2}}, \,\, 0<r_{M}<\infty.
	\end{equation}
	Denote $b(u_0,r_M)$ as the circle of radius $r_M$ centered at $u_0$. Given there are $m$ BSs in $b(u_0,r_M)$, the distribution of these $m$ BSs is a homogeneous binomial point process, and hence the conditional joint PDF is
	$f_{\!\!\left.R_1,R_2,\cdots\!, R_m \right| R_M}\! (r_1,r_2 \cdots\!, r_m) \!=\! \prod_{i=1}^{m} \frac {2r_i} {r_M^2}, \,\, 0 \!<\! r_i\! \leq\! r_M.$
%	\begin{equation}\label{equ: appendix Joint conditional PDF R1_RM}
%	f_{\!\!\left.R_1,R_2,\cdots\!,\! R_m \right| R_M}\! (r_1,r_2 \cdots\!, r_m) \!=\! \prod_{i=1}^{m} \frac {2r_i} {r_M^2}, \,\, 0 \!<\! r_i\! \leq\! r_M.
%	\end{equation}
	Combining with \eqref{equ: appendix PDF of R_M} completes the proof of Lemma \ref{lemma Joint pdf of BSs distances}.
	
\section{Proof of Theorem \ref{theorem: STP in NC-JT} }\label{appendix: Proof STP in NC-JT}
	For NC-JT, $s=\text{noCSI}$. According to \eqref{equ: q^s 2}, we first calculate the probability mass function $\operatorname{Pr} \left[ \left| \mathcal{C}_{n} \right| =m\right], m=0,1,\cdots,M$. Each BS stores file $n$ randomly with probability $T_n$ according to the probabilistic content placement strategy. Besides, there are at most $M$ BSs in $\mathcal{C}_{n}$, therefore, $\left| \mathcal{C}_{n} \right|$ follows a binomial distribution with parameters $M$ and $T_n$, i.e., $\operatorname{Pr}\left[\left|\mathcal{C}_{n}\right|=m\right] = \dbinom{M}{m} T_{n}^{m} \left( 1 - T_{n} \right) ^ {M-m}$. Next, we calculate $q_{n,0}(T_n)$ and $q_{c,m} ^{\text{noCSI}} $ separately. 
	
	To calculate $q_{n,0}(T_n)$, we rewrite the interference in \eqref{equ: SINR_n} as $I = I_n+I_{-n}$, where $I_{n} \triangleq \sum_{x \in \Phi_{b,n} \backslash \{x_0\}} \|x\|^{-\alpha} \left|h_{x}\right|^{2} $ and $I_{-n} \triangleq \sum_{x \in \Phi_{b,-n} \backslash \mathcal{C}} \|x\|^{-\alpha} \left|h_{x}\right|^{2} $, and $x_0$ denotes the only serving BS of $u_0$, the distance between whom is $R_0$. We have $R_0\!\!>\!\!R_M$, the normalized received power is $ S^{\text{noCSI}}\!\! =\!\! R_0^{-\!\alpha}\! \left| h_{x_0} \right|^2 $. In this case, we have
	\begin{equation}\label{equ: appendix qn0}
	\begin{aligned}
	q_{n,0}(T_n)&\! =\! \operatorname{Pr}\left[\text{SIR}_{n} \geq \tau | \left| \mathcal{C}_{n} \right| = 0 \right]
	\\& \!= \!\int_0 ^\infty \int_0 ^{r_0} f_{\left.R_0,R_M\right|R_0>R_M} \left( T_n, r_0, r_M \right) 
	\\&\times \!\operatorname{Pr}\!\left[\frac{R_0^{-\alpha} \left| h_{x_0} \right|^2}{I_n+I_{-n}}\! \geq\! \tau\! \left.\! \right| \! R_0 = r_0, \!R_M \!= \!r_M\! \right] \!\mathrm{d}r_{\!M} \mathrm{d}r_0,
	\end{aligned}
	\end{equation}
	where $f_{\left.R_0,R_M\right|R_0>R_M} \left( T_n, r_0, r_M \right)$ is the conditional joint PDF of $R_0$ and $R_M$ conditioning on $R_0>R_M$. From \eqref{equ: appendix PDF of R_M}, we denote the PDF of the distance of the $i$-th nearest BS as $f_i(x,\lambda)$. Note that due to the silencing of BSs, the serving BS is the first-nearest BS in $\Phi_{b,n}$ and the first interfering BS is the $M$-th nearest BS in $\Phi_{b,-n}$. Considering the independence between $\Phi_{b,n}$ and $\Phi_{b,-n}$, we have
	\begin{equation}\label{equ: appendix f_R0RM}
	\begin{aligned}
	&f_{\left.R_0,R_M\right|R_0>R_M} \left( T_n, r_0, r_M \right) 
	\\&\,\,= \frac{ f_1(r_0, \lambda_{b}T_n) f_M(r_M, \lambda_{b}(1-T_n))} {\operatorname{Pr}(R_0>R_M)}
	\\&\,\,= \frac{4 \pi ^{ M+1 } \lambda_{b} ^{ M+1 } T_{n} r_0 r_M ^{2 M-1} } { \Gamma(M) }
	\\ &\,\,\quad\times\exp \left(-\pi \lambda_{b} T_{n} r_0^{2} -\pi \lambda_{b} \left( 1-T_{n} \right) r_M^{2} \right),
	\end{aligned}
	\end{equation}
	where $\operatorname{Pr}(R_0>R_M) = (1-T_n)^M$.	Next, we have
	\newcounter{TempC3}
	\setcounter{TempC3}{\value{equation}}
	\setcounter{equation}{27}
	\begin{figure*}[t]	
		\begin{equation}\label{equ: appendix q_cm_noCSI }
		\begin{aligned}
		q_{c,m} ^{\text{noCSI}} \!\!= &\operatorname{Pr}\left[\text{SIR}_{n} \geq \tau | \left| \mathcal{C}_{n} \right| = m \right]
		\\ =&\operatorname{Pr}\left[\text{SIR}_{n} \geq \tau,x_M \notin \mathcal{C}_{n} \left| \right. \left| \mathcal{C}_{n} \right| = m \right] 
		+\operatorname{Pr}\left[\text{SIR}_{n} \geq \tau, x_M \in \mathcal{C}_{n} \left| \right. \left| \mathcal{C}_{n} \right| = m \right]
		\\ =&\! \underbrace { \operatorname{Pr}\! \left[S^{\text{noCSI}}\!/\!I \!\geq\! \tau\! \left| \right. \!x_M \!\notin\! \mathcal{C}_{n},\! \left| \mathcal{C}_{n} \right| \!=\! m \right] } _{ \triangleq q^{\text{noCSI}} _{c,m,1}}\! \operatorname{Pr}\! \left[ x_M \!\notin\! \mathcal{C}_{n} \left|\right. \left| \mathcal{C}_{n} \right|\! =\! m \right] \!+\! \underbrace {\operatorname{Pr}\!\left[\! S^{\text{noCSI}} \!/\!I\! \geq\! \tau\! \left| \right.\! x_M \!\in \!\mathcal{C}_{n},\!\left| \mathcal{C}_{n} \right|\! = \!m \right] }_ { \triangleq q^{\text{noCSI}} _{c,m , 2}}\!\operatorname{Pr}\! \left[ x_M \!\in\! \mathcal{C}_{n} \left|\right. \left| \mathcal{C}_{n} \right|\! = \! m \right].
		\end{aligned}
		\end{equation}	
		\hrulefill
	\end{figure*}
	\setcounter{equation}{\value{TempC3}}
	\begin{equation}
	\begin{aligned}
	\operatorname{Pr}&\left[\frac{R_0^{-\alpha} \left| h_{x_0} \right|^2}{I_n+I_{-n}} \geq \tau \left. \right| R_0 = r_0, R_M = r_M \right]
	\\&= \mathbb{E}_{I_n,I_{-n}}\left[ \operatorname{Pr} \left[ \left| h_{x_0} \right|^2 \geq \tau r_0^{\alpha} \left( I_n + I_{-n} \right) \right] \right]
	\\& \overset{(a)}{=} \mathbb{E}_{I_n,I_{-n}}\left[ \exp \left( - \tau r_0^{\alpha} \left( I_n + I_{-n} \right) \right) \right]
	\\&\overset{(b)}{=} \underbrace{ \mathbb{E}_{I_n}\left[ \exp \left( - s I_n \right) \right] }_{\triangleq \mathcal{L}_{I_{n}} \left. \left( s,r_0,r_M \right) \right|_{s=\tau r_0^{\alpha}} } \underbrace{\mathbb{E}_{I_{-n}}\left[ \exp \left( - s I_{-n} \right) \right] }_{\triangleq \mathcal{L}_{I_{-n}} \left. \left( s,r_0,r_M \right) \right|_{s=\tau r_0^{\alpha}} }
	\end{aligned}
	\end{equation}
	where (a) is due to $\left|h_{x_0}\right|^{2}\stackrel{d}{\sim} $ Exp(1)$ $, (b) is due to the independence of the homogeneous PPPs. $\mathcal{L}_{I_{n}} \left( s,r_0,r_M \right) $ and $\mathcal{L}_{I_{-n}} \left( s,r_0,r_M \right)$ represent the Laplace transforms of the interference $I_{n}$ and $I_{-n}$, respectively, and can be calculated as follows:
	\begin{equation}\label{equ: appendix L_In}
	\begin{aligned}
	&\mathcal{L}_{I_{n}} \left( s,r_0,r_M \right) \left.\right|_{ s = \tau r_0^{\alpha}}
	\\&\,\,=\mathbb{E}_{\Phi_{b,n}, \left| h_x \right| } \!\!\left[ \exp\left( -s \sum_{x \in \Phi_{b,n}\backslash \{ x_0 \} } \| x \| ^{-\alpha} \left| h_x \right| ^2\right) \right]
	\\& \,\,\overset{(a)}{=} \mathbb{E}_{\Phi_{b,n}}\!\! \left[ \prod_{x \in \Phi_{b,n} \backslash \{ x_0 \} } \!\!\!\mathbb{E}_{\left| h_x\right|}\!\! \left[ \exp \left( -s \| x \| ^{-\alpha} \left| h_x \right| ^2\right) \right] \right]
	\\ &\,\,\overset{(b)}{=}\mathbb{E}_{\Phi_{b,n}}\!\! \left[ \prod_{x \in \Phi_{b,n} \backslash \{ x_0 \} } \frac{1}{ 1 + s \| x \|^{-\alpha} }\right]
	\\ &\,\,\overset{ (c) } {=} \exp\left( -2\pi\lambda_b T_n \int_{r_0} ^{\infty} \left( 1-\frac{1}{1+s r^{-\alpha}} \right) r\mathrm{d}r \right)
	\\&\,\,= \exp\!\left( -2 \pi \lambda_{b} T_n \frac{ r_0^2}{\alpha -2} \frac{s}{r _0 ^\alpha} F_G\left( \alpha, -\frac{s}{r _0 ^\alpha} \right)\right)
	\end{aligned}
	\end{equation}
	\begin{equation}\label{equ: appendix L_I-n}
	\begin{aligned}
	& \mathcal{L}_{I_{-n}}\!\! \left.\left( s,r_0,r_M \right) \right|_{ s = \tau r_0^{\alpha}}
	\\&\,\, = \mathbb{E}_{\Phi_{b,-n}} \left[ \prod_{x \in \Phi_{b,-n} \backslash \mathcal{C} } \mathbb{E}_{\left| h_x\right|} \exp \left( -s \| x \| ^{-\alpha} \left| h_x \right| ^2\right) \right]
	\\ &\,\,= \exp\left( -2 \pi \lambda_{b} \left( 1 - T_n \right) \frac{r_M^2}{\alpha -2} \frac{s}{r _M ^\alpha} F_G\left( \alpha, -\frac{s}{r _M ^\alpha} \right) \right)
	\end{aligned}
	\end{equation}
	where (a) is due to the independence of the channels; (b) follows from $\left|h_{x}\right|^{2} \stackrel{d} {\sim} $ Exp(1)$ $; (c) is from the probability generating functional for a PPP \cite{01Haenggi2012StochasticGF} and converting from Cartesian to polar coordinates. Substituting \eqref{equ: appendix f_R0RM}, \eqref{equ: appendix L_In} and \eqref{equ: appendix L_I-n} into \eqref{equ: appendix qn0} and using \eqref{equ: A} and the change of variable $ u = \pi\lambda_{b}T_n r^2 $, we can obtain $q_{n,0}(T_n)$.
	
	Next, we calculate $q_{c,m} ^{\text{noCSI}} $. Let $x_M$ be the $M$-th nearest BS in $\mathcal{C}$. We consider two cases: i) $x_M \notin \mathcal{C}_n$ and ii) $x_M \in \mathcal{C}_n$. Conditioning on $\left| \mathcal{C}_{n} \right| = m $, the normalized received power at $u_0$ is $ S^{\text{noCSI}} = \left| \sum_{i=1}^{m} \| x_i \| ^{-\alpha/2} h_{x_i} \right|^2 $, and the normalized interference is $I = \sum_{x \in \Phi_{b}^{c}} \|x\| ^{-\alpha}\left|h_{x}\right|^{2}$. Then, we have \eqref{equ: appendix q_cm_noCSI }.
	Due to the probabilistic content placement strategy, we have $\operatorname{Pr}\left[ x_M \notin \mathcal{C}_{n} \left|\right. \left| \mathcal{C}_{n} \right|= m \right] =1-\frac{m}{M}$ and $\operatorname{Pr}\left[ x_M \in \mathcal{C}_{n} \left|\right. \left| \mathcal{C}_{n} \right|= m \right] =\frac{m}{M},m=1,2,\cdots,M$. 
	As for $q^{\text{noCSI}} _{c,m,1}$, when $m=M$, all the BSs in $\mathcal{C}$ jointly serve $u_0$ and $x_M\notin \mathcal{C}_{n}$ can not happen, thus, we set $q^{\text{noCSI}} _{c,m,1}=0$ in this case. When $m<M$, we take a condition that $\mathbf{R} = \mathbf{r}$, whose definition is given in Lemma \ref{lemma Joint pdf of BSs distances}. We have:
	\setcounter{equation}{28}
	\begin{equation} \label{equ: appendix qnoCSIcm1R}
	\begin{aligned}
	q&^{\text{noCSI}}_{c,m,1,\mathbf{R}}\left(\mathbf{r}\right)
	\\&= \operatorname{Pr} \left[ S^{\text{noCSI}}/I \geq \tau \left| \right.\mathbf{R} = \mathbf{r}, x_M \notin \mathcal{C}_{n}, \left| \mathcal{C}_{n} \right| = m \right] 
	\\ &=\mathbb{E}_I \left[ \operatorname{Pr} \left[ S^{\text{noCSI}} \geq \tau I \left| \!\right.\mathbf{R} \!= \!\mathbf{r}, x_M \notin \mathcal{C}_{n}, \left| \mathcal{C}_{n} \right| = m \right] \right]
	\\ &\overset{ \text{(a)} } {=} \mathbb{E}_I \left[ \exp \left( - \frac{\tau I}{\sum_{i=1}^{m} r_i^{-\alpha}} \right) \right]
	\\&\triangleq \mathcal{L}_I \left. \left( s,\mathbf{r}\right) \right|_{s=\tau/ \sum_{i=1}^{m} r_i^{-\alpha}},
	\end{aligned}
	\end{equation}
	where (a) follows from that $ \left| \sum_{i=1}^{m} \| x_i \| ^{-\alpha/2} h_{x_i} \right|^2 \stackrel{d}{\sim} $ Exp$\left( \frac{1} {\sum_{i=1}^{m} r_i^{-\alpha} } \right) $.
	$ \mathcal{L}_I \left( s,\mathbf{r}\right) $ can be calculated similarly to \eqref{equ: appendix L_In}:
	\begin{equation} \label{equ: appendix L I}
	\begin{aligned}
	\left. \mathcal{L}_I \!\left( s,\!\mathbf{r}\right) \right|_{s \!=\!\frac{\tau}{\sum\limits_{i=1}^{m}\!\! r_i^{-\alpha}}} \!=\! \exp \! \left(\!\! -\!2\pi\!\lambda_{b} \frac{r_M^2}{\alpha\! - \!2} \frac{s}{r_M^\alpha} \!F_G \!\left(\! \alpha,\! -\frac{s}{r _M ^\alpha} \! \right)\!\!\right)\!.
	\end{aligned}	
	\end{equation}
	Now we calculate $q^{\text{noCSI}} _{c,m,1}$ by removing the condition $\mathbf{R} = \mathbf{r}$, whose joint PDF $f_{\mathbf{R}}(\mathbf{r})$ is given by Lemma \ref{lemma Joint pdf of BSs distances}. We have
	\begin{equation} \label{equ: appendix qcm1NoCSI integral}
	\begin{aligned}
	q^{\text{noCSI}}_{c,m,1}\!=\! \int\limits_{0}^{\infty}\! \int\limits_{0}^{r_M}\! \cdots\! \int\limits_{0}^{r_M} \!q^{\text{noCSI}}_{c,m,1,\mathbf{R}}\!\left(\mathbf{r}\right) \!f_{\mathbf{R}}\left(\mathbf{r}\right) \mathrm{d}r_1\cdots \mathrm{d}r_m \mathrm{d}r_M 
	\end{aligned}.
	\end{equation}
	By using \eqref{equ: A}, the changes of variables $u=\pi \lambda_{b} r_M^{2}$ and $t_i = \frac{r_i^2}{r_M^2}$, and the definition of $R_{m,1}(x,\beta)$ given in \eqref{equ: R_1}, we can obtain $q^{\text{noCSI}}_{c,m,1} =R_{m,1}(1,1) $ in \eqref{equ: appendix q_cm_noCSI }.
	
	As for $q^{\text{noCSI}}_{c,m,2}$, when $m=1$, the only serving BS is the $M$-th nearest BS. In this case, we can get $q^{\text{noCSI}}_{c,m,2,R_M}\left(r_M\right)$ by setting $s = \tau /r_M^{-\alpha}$ in \eqref{equ: appendix L I}, and the corresponding joint PDF in Lemma \ref{lemma Joint pdf of BSs distances} degenerates to $ f_{R_M}(r_M) = \frac{ 2(\pi\lambda_b)^M } {\Gamma(M)} r_M^{2M-1} e^{-\pi\lambda_b r_M^2 } $. The rest of the proof is similar to that of calculating $q^{\text{noCSI}} _{c,m,1}$. We omit the details due to page limitations. Combing the three parts introduced above, we prove Theorem \ref{theorem: STP in NC-JT}.
	
\section{Proof of Theorem \ref{theorem: upper and Appr of C-JT}}\label{appendix: Proof of STP in C-JT}
	We just give the proof of $ q_{c,m}^{\text{CSI},u}$, and the rest parts of $q^{\text{CSI},u} ( \mathbf{T} )$ can be derived by following the similar steps shown in Appendix \ref{appendix: Proof STP in NC-JT}. Similar to $q_{c,m}^{\text{noCSI}}$ given in \eqref{equ: appendix q_cm_noCSI }, we have $q_{c,m}^{\text{CSI},u} \triangleq \left(1-\frac{m}{M} \right) q_{c,m,1}^{\text{CSI},u} + \frac{m}{M} q_{c,m,2} ^{\text{CSI},u}$, and to calculate $q_{c,m,1}^{\text{CSI},u}$, we take a condition that $\mathbf{R} = \mathbf{r}$. When $m =1,2,\cdots,M-1$, we have
	\begin{equation} \label{equ: appendix qm1RCSI }
	\begin{aligned}
	&q_{c,m,1,\mathbf{R}}^{\text{CSI},u}\left(\mathbf{r}\right)
	\\&=\!\operatorname{Pr}\left[ S^{\text{CSI}}/I \geq \tau \left| \right.\mathbf{R} = \mathbf{r}, x_M \notin \mathcal{C}_{n}, \left| \mathcal{C}_{n} \right| = m \right]
	\\ &=\!\mathbb{E}_I\! \Bigg[\!\! \operatorname{Pr}\! \Bigg[\!\!\left(\! \sum_{i=1}^{m} r_i ^{-\!\frac{\alpha}{2}}\! \!\left| h_{x_i} \!\right| \!\right)^2\!\!\!\! \geq\! \tau \!I \!\left| \right.\! \mathbf{R} \!=\! \mathbf{r}, b_M\!\! \notin\! \mathcal{C}_{n},\! \left| \mathcal{C}_{n} \right|\! = \!m \Bigg] \Bigg]
	\\ & \!\overset{(a)}{\leq}\! \mathbb{E}_I \! \Bigg[\! \!\operatorname{Pr}\! \Bigg[ \!\sum_{i=1}^{m} \!\left| h_{x_i} \right| ^2\! \geq\!\! \frac{\tau I}{\omega} \left| \!\right. \mathbf{R}\! =\! \mathbf{r}, b_M \!\!\notin \!\mathcal{C}_{n},\! \left| \mathcal{C}_{n} \right| = m \Bigg] \Bigg]
	\\ &\overset{(b)}{=} \!\mathbb{E}_I\! \left[ 1 - \frac{\gamma \left(m, \frac{\tau I} {\omega} \right)}{\Gamma(m)}\right]
	\!\overset{(c)}{\leq} \!1 \!-\! \mathbb{E}_I \left[ \left(1-e^{-\beta \frac{\tau I} {\omega} }\right)^{m} \!\right]
	\\ & \overset{(d)}{=} \sum_{j=1}^{m} (-1)^{j+1} \dbinom{m}{j} \underbrace {\mathbb{E}_I \left[\exp{\left( -j \beta \frac{\tau I} {\omega} \right) } \right] }_{\triangleq \left. \mathcal{L}_I \left( s,\mathbf{r}\right) \right|_{s= j \beta \frac{\tau}{ \omega }}},
	\end{aligned}
	\end{equation}
	here, $\omega = \sum _{i=1} ^{m} r_i ^{-\alpha}$, and $ \left| h_{x_i} \right| $ are i.i.d. Rayleigh RVs with scale parameter $\sigma = \sqrt{2}/2$, and hence $\sum_{i=1}^{m} \left| h_{x_i} \right| ^2 \overset{d}{\sim} Gamma(m,1)$; (a) follows from Lemma \ref{lemma: L Bound on WSR RVs}; (b) follows from the CDF of gamma distribution; (c) follows from the lower bound in Lemma \ref{lemma: Bound on IGF} and $\beta = \Gamma(m+1)^{-1 / m}$; (d) is from the binomial theorem.
	
	Similar to \eqref{equ: appendix qnoCSIcm1R} and \eqref{equ: appendix L I}, we have
	\begin{equation}
	\begin{aligned}
	\mathcal{L}_I \!\left( s,\mathbf{r}\right) \left.\!\!\right|_{s= j \beta \frac{\tau}{ \omega }}\! = \!\exp\!\left(\!\! -2\pi\lambda_{b} \frac{r_M^2}{\alpha -2} \frac{s}{r_M^\alpha} \!F_G \!\left(\! \alpha,\! -\frac{s}{r _M ^\alpha} \! \right) \!\right)
	\end{aligned}
	\end{equation}
	Then, $q^{\text{CSI},u} _{c,m,1}$ can be obtained by removing the condition of $q^{\text{CSI},u}_{c,m,1,\mathbf{R}}\left(\mathbf{r}\right)$ on $\mathbf{R} = \mathbf{r}$ as \eqref{equ: appendix qcm1NoCSI integral}. When $m = M$, $q_{c,m,1}^{\text{CSI},u} = 0$ as illustrated in Appendix \ref{appendix: Proof STP in NC-JT}. The proof of $q^{\text{CSI},u} _{c,m,2}$ is similar to that of $q^{\text{CSI},u} _{c,m,1}$, and we omit the details.
	
	To obtain the approximation of $q^{\text{CSI}}(\mathbf{T})$, i.e. $q^{\text{CSI},a} ( \mathbf{T} )$, we just take the upper bound in Lemma \ref{lemma: Bound on IGF}, which means substituting $\beta = \Gamma(m+1)^{-1 / m}$ with $\beta = 1$, and the remaining proof is the same as that of $q^{\text{CSI},u} ( \mathbf{T} )$. The tightness of this approximation will be demonstrated by comparing with simulation results in the following part.
\section{Proof of Theorem \ref{theorem: Solution of P1}} \label{appendix: Proof Optimal solution of P1}
	We just give the proof for NC-JT, i.e., $g$ is set as ``noCSI''. The Lagrange function of Problem \ref{problem: 1} is
	\begin{equation}
	\begin{aligned}
	L(\mathbf{T}, \boldsymbol{\lambda}, \boldsymbol{\eta}, \nu) =& -\sum_{n \in \mathcal{N}} a_{n} q_{n}^{\text{noCSI}}\left(T_{n}\right) - \sum_{n \in \mathcal{N}} \lambda_{n} T_{n} \\&+\!\!\sum_{n \in \mathcal{N}} \eta_{n}\! \left( T_{n}-1 \right)\! +\!\nu\! \left(\sum_{n \in \mathcal{N}} T_{n} - K\! \right)
	\end{aligned}	
	\end{equation}
	where $ \boldsymbol{\lambda} \triangleq \left( \lambda_{n} \right) _ { n \in \mathcal{N}}$ and $\boldsymbol{\eta} \triangleq \left( \eta_{n} \right) _{n \in \mathcal{N}}$ are the Lagrange multipliers associated with the constraints \eqref{equ: constraint of T 1}. $\nu$ is the Lagrange multiplier associated with the constraint \eqref{equ: constraint of T 2}. Thus, we have 
	\begin{equation}
	\frac{\partial L ( \mathbf{T}, \boldsymbol {\lambda}, \boldsymbol{\eta}, \nu)}{\partial T_n} = 	-a_n D^{\text{noCSI}}_n (T_n) - \lambda_n + \eta_n + \nu.
	\end{equation}
	Then, the KKT conditions can be written as
	\begin{eqnarray}
	0 \leq T_{n}^\star \leq 1,\,\,\sum\nolimits_{n \in \mathcal{N}} T_{n}^\star &=& K, \,\,\, \forall n \in \mathcal{N},\label{equ: KKT 1}\\
	\lambda_n \geq 0,\,\, \eta_n &\geq& 0, \quad \forall n \in \mathcal{N},\label{equ: KKT 2}\\
	\lambda_n T_n^\star = 0,\,\, \eta_n (T_n^\star-1) &=& 0, \quad \forall n \in \mathcal{N},\label{equ: KKT 3} \\
	-a_n D^{\text{noCSI}}_n (T_n^\star) - \lambda_n + \eta_n + \nu &=& 0,\quad \forall n \in \mathcal{N}. \label{equ: KKT 4}
	\end{eqnarray}
	We rewrite the condition \eqref{equ: KKT 4} as $\eta_n =a_n D^{\text{noCSI}}_n (T_n^\star) + \lambda_n - \nu $. Thus, we have: (a) if $\nu > a_n D^{\text{noCSI}}_n (T_n^\star)$, to satisfy the dual constraint $\eta_n \geq 0$, we need $\lambda_n>0$, according to the complementary slackness $\lambda_n T_n^\star = 0$, we have $T_n^\star = 0$; (b) if $\nu < a_n D^{\text{noCSI}}_n (T_n^\star)$, considering the dual constraints \eqref{equ: KKT 2} and the complementary slackness $\eta_n (T_n^\star - 1)=0$, we have $T_n^\star\! = \!1$; (c) if $a_n D^{\text{noCSI}}_n (1) \!\leq \!\nu \!\leq \! a_n D^{\text{noCSI}}_n (0)$, satisfying \eqref{equ: KKT 2} and \eqref{equ: KKT 3} leads to $\lambda_n = 0$ and $\eta_n = 0$, resulting in $\nu = a_n D^{\text{noCSI}}_n (T_n^\star)$. Combining with \eqref{equ: constraint of T 1} and \eqref{equ: constraint of T 2}, we proof Theorem \ref{theorem: Solution of P1}.
\section{Proof of Lemma \ref{lemma: Property of Optimal Solution for Problem 1 }}\label{appendix: Proof property of optimal solution for P1}
	We just give the proof in NC-JT, i.e., $g$ is set as ``noCSI''. Consider $m,n\in \mathbb{N}^{+}$, $m<n$ (i.e., $a_m>a_n$). $T_m^\star$ and $T_n^\star$ are the corresponding optimal caching probabilities of the file $m$ and $n$, respectively. To prove this lemma, we need to prove that $T_m^\star \geq T_n^\star$. There are three cases: a) If $T_n^\star=0$, $T_m^\star \geq T_n^\star$ is satisfied obviously. b) If $T_n^\star=1$, to ensure $T_m^\star \geq T_n^\star$, we need $T_m^\star=1$. From \eqref{equ: Optimal Solution of P1}, we have $a_n D^{\text{noCSI}}_n(1) > \nu$. Supposing that $T_m^\star \neq 1$, notice that $D^{\text{noCSI}}_m(T_m)$ decreases with $T_m$ since $q_{m}(T_m)$ is a concave function of $T_m$ when $ q_{c,m+1} - q_{c,m} \leq q_{c,m} - q_{c,m-1} $ for all $m = 2,\cdots,M-1$, we have $D^{\text{noCSI}}_m(T_m^\star) > D^{\text{noCSI}}_m(1)$, then we have $a_m D^{\text{noCSI}}_m(T_m^\star) > a_m D^{\text{noCSI}}_m(1) > a_n D^{\text{noCSI}}_m(1) = a_n D^{\text{noCSI}}_n(1) >\nu$. According to \eqref{equ: Optimal Solution of P1}, we have $T_m^\star=1$, which conflicts with $T_m^\star \neq 1$. Thus, we have $T_m^\star=1$, and hence, $T_m^\star \geq T_n^\star$. c) If $T_m^\star = x\left(T_{m}^\star, a_{m}, \nu \right)$, $\nu = a_m D^{\text{noCSI}}_m(T_m^\star)$ holds. Supposing that $T_m^\star < T_n^\star$, we have $D^{\text{noCSI}}_m(T_m^\star) > D^{\text{noCSI}}_n(T_n^\star)$, and $a_n D^{\text{noCSI}}_n(T_n^\star) > a_m D^{\text{noCSI}}_m(T_m^\star) =\nu$, yielding $T_m = 1$, which conflicts with $T_m^\star < T_n^\star$. Thus, we have $T_m^\star \geq T_n^\star$. In summary, we can prove Lemma \ref{lemma: Property of Optimal Solution for Problem 1 }.
	
	%\section*{Acknowledgment}
	%This work was supported by
	
	\ifCLASSOPTIONcaptionsoff
	\newpage
	\fi
	\bibliographystyle{IEEEtran}
	\bibliography{IEEEabrv,mylib_abrv}

\end{document}